\documentclass[prd,twocolumn,showpacs,floatfix,amsmath,nofootinbib,amssymb,floatfix]{revtex4}
\usepackage{graphicx,color,dcolumn,booktabs,bm}
\usepackage{longtable,lscape}
\usepackage{txfonts}
\usepackage{overpic}
\usepackage{amssymb}
\usepackage{indentfirst}
\usepackage{feynmf}   
\usepackage{slashed}  
\usepackage{cases}
\usepackage{color}
\usepackage{multirow}
\usepackage{epstopdf}
\usepackage{enumerate}
\usepackage{graphicx,color,dcolumn,booktabs,bm}
\usepackage[colorlinks, citecolor=blue,anchorcolor=red,menucolor=red, linkcolor=red,filecolor=red,urlcolor=blue,frenchlinks=red]{hyperref}

\graphicspath{{Figures/}} %

\begin{document}

\title{
Are the $Y$ states around 4.6 GeV from $e^+e^-$ annihilation higher charmonia?
}
\author{Jun-Zhang Wang$^{1,2}$}\email{wangjzh2012@lzu.edu.cn}
\author{Ri-Qing Qian$^{1,2}$}\email{qianrq18@lzu.edu.cn}
\author{Xiang Liu$^{1,2}$\footnote{Corresponding author}}\email{xiangliu@lzu.edu.cn}
\author{Takayuki Matsuki$^{3,4}$}\email{matsuki@tokyo-kasei.ac.jp}
\affiliation{$^1$School of Physical Science and Technology, Lanzhou University, Lanzhou 730000, China\\
$^2$Research Center for Hadron and CSR Physics, Lanzhou University $\&$ Institute of Modern Physics of CAS, Lanzhou 730000, China
\\$^3$Tokyo Kasei University, 1-18-1 Kaga, Itabashi, Tokyo 173-8602, Japan
\\$^4$Theoretical Research Division, Nishina Center, RIKEN, Wako, Saitama 351-0198, Japan}

\date{\today}

\begin{abstract}
In this work, we present the mass spectrum of higher charmonia around and above 4.6 GeV by adopting the unquenched potential model. We perform a combined fit to the updated experimental data of $e^+e^- \to \psi(2S)\pi^+\pi^-$ and $e^+e^-\to \Lambda_c\bar{\Lambda}_c$. To understand the ``platform" structure observed in the range of $4.57
\sim 4.60$ GeV existing in the $\Lambda_c\bar{\Lambda}_c$ invariant mass spectrum of
$e^+e^-\to \Lambda_c\bar{\Lambda}_c$ of BESIII, we introduce two resonances in this combined fit to the $e^+e^- \to \psi(2S)\pi^+\pi^-$ and $e^+e^-\to \Lambda_c\bar{\Lambda}_c$, which have resonance parameters, $m_{Y_1}=4585\pm2$ MeV, $\Gamma_{Y_1}=29.8\pm8.0$ MeV, $m_{Y_2}=4676\pm7$ MeV, and $\Gamma_{Y_2}=85.7\pm15.0$ MeV.
Furthermore, combining with our theoretical results, we indicate that the two charmonium-like $Y$ states can be due to two higher charmonia, which are mixtures of $6S$ and $5D$ $c\bar{c}$ states. Their two-body open-charm decay behaviors are given. Under the same framework, our analysis of the data of $e^+e^-\to D_{s}^+D_{s1}(2536)^-$ recently released by Belle supports to introduce these two higher charmonia around 4.6 GeV. Additionally, we predict the masses and two-body open-charm decays of six higher charmonia $\psi(nS)$ and $\psi(mD)$ with $n=7,8,9$ and $m=6,7,8$ above 4.6 GeV. Search for these higher charmonia will be an interesting issue for the running BESIII and BelleII, and even the possible Super Tau-Charm Factory discussed in China.

\end{abstract}
 \maketitle

\section{\label{sec1}Introduction}

Since the observation of $J/\psi$ in 1974 \cite{Aubert:1974js,Augustin:1974xw}, construction of charmonium family has become an interesting issue of the study in hadron spectroscopy, which 
provides a valuable hint to deepen our understanding of non-perturbative behavior of quantum chromodynamics (QCD). Especially, discovery of a series of  charmonium-like $XYZ$ states in the past 16 years not only inspires extensive discussions on exotic hadrons, but also provides a good chance to explore higher charmonia, which is not an easy task. More theoretical and experimental efforts should be done in the near future.

\begin{figure}[htbp]
\centering
\includegraphics[width=6cm,keepaspectratio]{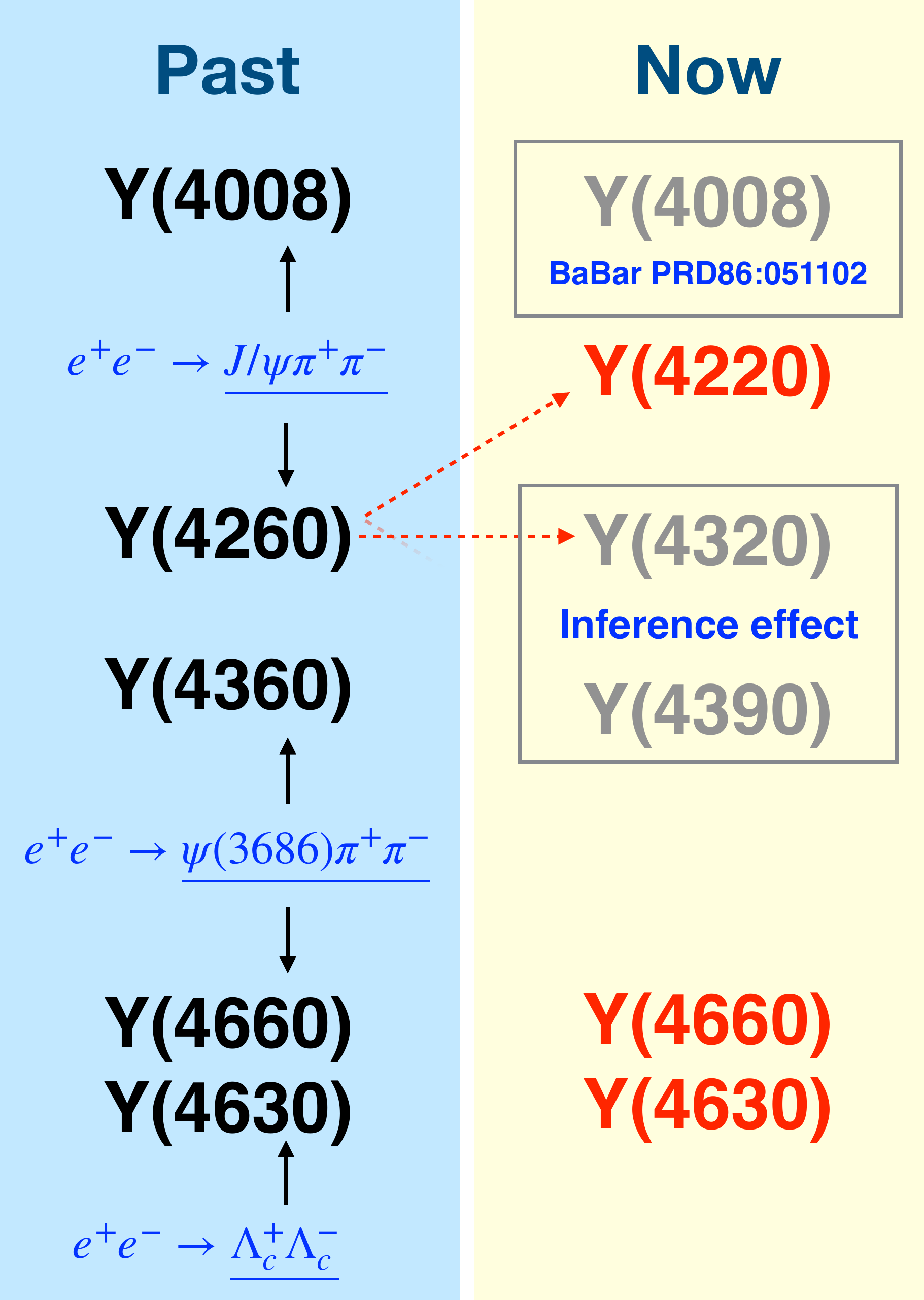}
\caption{The status of charmonium-like $Y$ states before 2017 and the present status of the $Y$ states after 2017. Since $Y(4008)$ was not confirmed by BaBar \cite{Lees:2012cn}, $Y(4008)$ is marked in grey. As for $Y(4320)$ and $Y(4390)$, the Lanzhou group indicated that they can be due to the Fano-like interference from $\psi(4160)$, $\psi(4415)$, and continuum contribution \cite{Chen:2017uof}. After considering these studies, there remain three charmonium-like $Y$ states, $Y(4220)$, $Y(4630)$, and $Y(4660)$, which are marked in red here.}
\label{nearbythresholdstates1}
\end{figure}

Among the observed $XYZ$ states, the $Y$ states from $e^+e^-$ annihilation form a special group. As once super star, $Y(4260)$ was firstly reported by BaBar in $e^+e^-\to J/\psi \pi^+\pi^-$ \cite{Aubert:2005rm}. Then, $Y(4360)$ was announced by Belle when analyzing $e^+e^-\to \psi(3686) \pi^+\pi^-$ \cite{Wang:2007ea}.
Up to date, more $Y$ states have been discovered, especially with the BESIII progress in 2017 \cite{Ablikim:2016qzw,BESIII:2016adj,Ablikim:2017oaf}. In Fig. \ref{nearbythresholdstates1}, we briefly summarize their status.

It is obvious that the readers must be confused 
when facing so many $Y$ states. It has become a crucial research topic focused by theorists (see review articles \cite{Chen:2016qju,Liu:2019zoy,Brambilla:2019esw} for a comprehensive progress) how to understand these novel phenomena. Since 2014, the Lanzhou group has proposed a Fano-like interference picture to reproduce $Y(4260)$ and $Y(4360)$, and further indicated that a narrow structure around 4.2 GeV as a charmonium should exist by the $c\bar{c}$ mass spectrum analysis \cite{Chen:2015bft,He:2014xna,Chen:2015bma}. The more precise BESIII data from the $e^+e^-\to \chi_{c0}\omega$ \cite{Ablikim:2014qwy}, $e^+e^-\to J/\psi\pi^+\pi^-$ \cite{Ablikim:2016qzw}, $e^+e^-\to h_c\pi^+\pi^-$\cite{BESIII:2016adj} and $e^+e^-\to \psi(3686)\pi^+\pi^-$ \cite{Ablikim:2017oaf} indeed show the existence of this narrow structure, which is collectively refered to as $Y(4220)$. With the updated data of charmonium-like $Y$ states, the Lanzhou group also proposed a $4S$-$3D$ mixing scheme to decode the nature of $Y(4220)$ and predict the $\psi(4380)$ as a partner of $Y(4220)$ \cite{Wang:2019mhs}. In addition, $\psi(4415)$ can be assigned to the $J/\psi$ family under a $5S$-$4D$ mixing scheme, and its partner $\psi(4500)$ was predicted. In fact, the theoretical work from the Lanzhou group provides a unified framework to settle down the observed $Y$ states from $e^+e^-$ annihilation \cite{Wang:2019mhs}. In Fig. \ref{nearbythresholdstates1}, we further give comparison of the past charmonium-like $Y$ states and three remaining $Y$ states when considering our unified framework and experimental progress.

There exist only two $Y$ states around 4.6 GeV. {The first observed state is $Y(4660)$, which was reported in the hidden-charm process of $e^+e^- \to \psi(2S)\pi^+\pi^-$ \cite{Wang:2007ea}. In the past years, theoretical explanations of the $Y(4660)$ state were 
 exotic hadron configurations like $f_0(980)\psi(2S)$ molecular state \cite{Guo:2008zg}, tetraquark state with diquark-antidiquark  [$cq$][$\bar{c}\bar{q}$] and [$cs$][$\bar{c}\bar{s}$] type \cite{Ebert:2008kb,Albuquerque:2008up,Albuquerque:2011ix,Chen:2010ze,Zhang:2010mw,Sundu:2018toi,Wang:2019iaa,Wang:2018rfw}, and hadro-charmonium of compact charmonium resonance inside an excited state of light hadronic matter \cite{Dubynskiy:2008mq}. Another $Y(4630)$ was observed in the process $e^+e^- \to \Lambda_c\bar{\Lambda}_c $ by Belle \cite{Pakhlova:2008vn}. Because it is near the production threshold of $\Lambda_c\bar{\Lambda}_c $, the $Y(4630)$ is considered to be a possible candidate of $\Lambda_{c}\bar{\Lambda}_{c}$ bound state \cite{Lee:2011rka} or charmed baryonium \cite{Cotugno:2009ys}. Although several exotic hadron configurations are very popular to explain these charmoniumlike $Y$ states. Due to our previous success in understanding the $Y$ states with lower mass in a unified framework \cite{Wang:2019mhs}, it is a natural conjecture that two $Y$ states around 4.6 GeV can be treated as higher charmonia. This conjecture directly motivates us to carry out the present work. }

In fact, the present work can be a sister paper of our former work \cite{Wang:2019mhs}. In this work, we will extend our calculation of charmonium mass spectrum to 4.6 GeV by adopting the same approach as that in Ref. \cite{Wang:2019mhs}. By combining with the experimental data, we will illustrate the possibility of assigning two $Y$ states around 4.6 GeV to the $J/\psi$ family, which we find can be due to the same source by performing a combined fit to $e^+e^-\to \psi(3686)\pi^+\pi^-$ and $e^+e^-\to \Lambda_c\bar{\Lambda}_c$ data. Additionally, we will predict other several vector charmonia above 4.6 GeV and discuss possible search plans.

On 5 November of 2019, we notice the Belle's new result of $e^+e^-\to D_s^+ D_{s1}(2536)^-$ \cite{Jia:2019gfe}, where a vector charmonium-like $Y$ state with mass $4625.9^{+6.2}_{-6.0}({\rm stat.})\pm0.4({\rm syst.})$ MeV and width $49.8^{+13.9}_{-11.5}({\rm stat.})\pm4.0({\rm syst.})$ MeV was reported. If treating it as the same state as either of $Y(4630)$ or $Y(4660)$, the possibility of assigning these $Y$ states around 4.6 GeV as higher charmonia is enforced, which we also discuss in this work.

This paper is organized as follows. Firstly, we study mass behaviour of vector charmonia around 4.6 GeV in Section \ref{sec:section1}. In Section \ref{section:fit}, the possibilities of two charmoniumlike $Y$ states around 4.6 GeV assigned to $J/\psi$ family are discussed by a combined analysis from theoretical results and experimental data of $e^+e^-\to\psi(2S)\pi^+\pi^-$ and $e^+e^-\to\Lambda_c\bar{\Lambda}_c$. Furthermore, mass spectrum and decay behaviours of higher charmonium states above 4.6 GeV are predicted in Section \ref{sec-higherccbar}. Finally, this paper ends with summary in Section \ref{sec-summary}.

\section{\label{sec:section1}Mass behaviors of vector charmonia around 4.6 GeV}


\subsection{A review of the unquenched potential model}
The interaction between quark and antiquark can be described by the following Hamiltonian \cite{Godfrey:1985xj}
	\begin{equation} \tilde{H}=(p^2+m_1^2)^{1/2}+(p^2+m_2^2)^{1/2}+V_{\mathrm{eff}}(\boldsymbol{p},\boldsymbol{r}),
	\end{equation}
where
	$V_{\mathrm{eff}}(\boldsymbol{p},\boldsymbol{r})=\tilde{H}^{\mathrm{conf}}+\tilde{H}^{\mathrm{hyp}}+\tilde{H}^{\mathrm{so}}$
is the effective potential of the $q\bar{q}$ interaction.


In the nonrelativistic limit, $V_{\mathrm{eff}}(\boldsymbol{p},\boldsymbol{r})$ is transformed into the standard nonrelativistic potential $V_{\mathrm{eff}}(r)$:
\begin{equation}
		V_{\mathrm{eff}}(r)=H^{\mathrm{conf}}+H^{\mathrm{hyp}}+H^{\mathrm{so}},
  \label{eq:eff}
\end{equation}
with
	\begin{equation}
		H^{\mathrm{conf}}=c+\frac{b(1-e^{-\mu r})}{\mu}+\frac{\alpha_s(r)}{r}\boldsymbol{F}_1\cdot\boldsymbol{F}_2,
\label{eq:potential}
	\end{equation}
where $H^{\mathrm{conf}}$ is the spin-independent potential and contains a constant term, a screened confining potential, and a one-gluon exchange potential. The screened potential depicts the unquenched effects from quark and antiquark pairs created from vacuum, where $\mu$ is a phenomenological parameter and reflects on the strength of unquenched effects. When $\mu$ goes to zero, the screened potential behaves like a linear confining potential $br$. The subscripts 1 and 2 denote quark and antiquark, respectively. The second term of Eq.~(\ref{eq:eff}) is the color-hyperfine interaction, i.e.,
	\begin{eqnarray}
		H^{\mathrm{hyp}}&=&-\frac{\alpha_s(r)}{m_1m_2}
		\Big[\frac{8\pi}{3}\boldsymbol{S}_1\cdot\boldsymbol{S}_2\delta^3(\boldsymbol{r})\nonumber\\
		&&+\frac{1}{r^3}\left(\frac{3\boldsymbol{S}_1\cdot\boldsymbol{r}\boldsymbol{S}_2\cdot\boldsymbol{r}}{r^2}-\boldsymbol{S}_1\cdot\boldsymbol{S}_2\right)\Big]
		\boldsymbol{F}_1\cdot\boldsymbol{F}_2, \label{eq:hyp}
	\end{eqnarray}
The third term  of Eq.~(\ref{eq:eff}) is the spin-orbit interaction
	\begin{equation}
		H^{so}=H^{so(\mathrm{cm})}+H^{so(\mathrm{tp})}.
	\end{equation}
Here, $H^{so(\mathrm{cm})}$ is the color-magnetic term and $H^{so(\mathrm{tp})}$ is the Thomas-precession term, i.e.,
	\begin{equation}
		H^{\mathrm{so}(\mathrm{cm})}=-\frac{\alpha_s(r)}{r^3}\left(\frac{1}{m_1}+\frac{1}{m_2}\right)\left(\frac{\boldsymbol{S}_1}{m_1}
+\frac{\boldsymbol{S}_2}{m_2}\right)\cdot\boldsymbol{L}(\boldsymbol{F}_1\cdot\boldsymbol{F}_2),
 \label{eq:socm}
	\end{equation}
	\begin{equation}
		H^{\mathrm{so}(\mathrm{tp})}=\frac{-1}{2r}\frac{H^{\mathrm{conf}}}{\partial r}\left(\frac{\boldsymbol{S}_1}{m_1}+\frac{\boldsymbol{S}_2}{m_2}\right)\cdot\boldsymbol{L}. \label{eq:sotp}
	\end{equation}

The relativistic corrections are introduced in two ways, the smearing transformation and momentum-dependent factors. The screened confining potential $S(r)=\frac{b(1-e^{-\mu r})}{\mu}+c$ \cite{Ding:1995he} and one-gluon exchange potential $G(r)=-4\alpha_s(r)/(3r)$ are smeared by \cite{Godfrey:1985xj}
\begin{equation}
\tilde{S}(r)/\tilde{G}(r)=\int d^3\boldsymbol{r}'\rho(\boldsymbol{r}-\boldsymbol{r}')S(r')/G(r'),
\end{equation}
where
\begin{equation}
\rho(\boldsymbol{r}-\boldsymbol{r}')=\frac{\sigma^3}{\pi^{3/2}}\mathrm{exp}\left[-\sigma^2(\boldsymbol{r}-\boldsymbol{r}')^2\right]
\end{equation}
is the smearing function.
The momentum dependent factors are introduced as \cite{Godfrey:1985xj}
\begin{equation}
    \tilde{G}(r)\to\left(1+\frac{p^2}{E_cE_{\bar{c}}}\right)^{1/2}\tilde{G}(r)\left(1+\frac{p^2}{E_cE_{\bar{c}}}\right)^{1/2},
\end{equation}
\begin{equation}
    \tilde{V}_i(r)\rightarrow \left(\frac{m_cm_{\bar{c}}}{E_cE_{\bar{c}}}\right)^{1/2+\epsilon_i}\tilde{V}_i(r)\left(\frac{m_cm_{\bar{c}}}{E_cE_{\bar{c}}}\right)^{1/2+\epsilon_i},
\end{equation}
where $E_c=\left(p^2+m^2_c\right)^{1/2}$, $\epsilon_i$ coresponds to different types of interaction, and $\tilde{V}_i(r)$ are potentials included in Eqs. (\ref{eq:hyp}, \ref{eq:socm}, \ref{eq:sotp}).
The details of unquenched potential model can be found in Ref. \cite{Wang:2019mhs}.

\subsection{charmonium spectrum results}
As stated in Ref. \cite{Li:2009ad}, the screened interaction in the unquenched quark model is partly equivalent to inclusion of the coupled channel effect. However, the screened potential is still a phenomenological treatment and related to non-perturbative behavior of QCD. So, we cannot directly know the parameter $\mu$. In general, the screening effects usually largely change the properties of high excited states, and hence the determination of the screened parameter $\mu$ must rely on experimental data of high excited charmonia.

In our previous work \cite{Wang:2019mhs}, we have studied the charmonium mass spectrum below 4.5 GeV by treating $Y(4220)$ as a scaling point of the screened parameter $\mu$. Based on this, we found that all observed charmoniumlike $Y$ states between 4.0 and 4.4 GeV together with $\psi(4415)$  can be explained very well in a unified framework. However,
there will be some problems when continuing to extend theoretical predictions to higher charmonia around 4.6 GeV. In fact, if we use a relatively low scaling point, the predictions for higher charmonia away from the scaling point will become uncertain in the unquenched potential model. The reason for this is that their mass behaviors are more sensitive to a screening parameter $\mu$ than the low-lying charmonia. Thus, in order to describe the charmonium spectrum around 4.6 GeV, a higher scaling point is necessary. In the following, we will carefully illustrate this point.

In Table \ref{table:mu}, we list sixteen charmonia so far established, where $Y(4220)$ and $\psi(4415)$ are treated as the scaling points of our unquenched potential model and are assigned to charmonia $\psi(4S)$ and $\psi(5S)$, respectively. In order to obtain the best agreement with the experimental mass of established charmonia, we mainly adjust three parameters $\mu$, $b$, and $c$ of the screened confining potential in Eq.~(\ref{eq:potential}). Here, we define a $\chi^2$ value to describe our theoretical deviation from experimental data, which is
	\begin{equation}
		\chi^2=\sum\left(\frac{M_{Exp}-M_{Th}}{M_{Er}}\right)^2,
	\end{equation}
where $M_{Exp}$, $M_{Th}$ and $M_{Er}$ are experimental value, theoretical value and error, respectively. An overall $M_{Er}$ = 5.0 MeV is chosen for convenience. From Table \ref{table:mu}, we can see that the established charmonium spectrum can be described well when the parameter $\mu$ is located in the range of $0.11\sim0.15$, where the masses of charmonia below 4.5 GeV have no obvious difference with different screened parameters. Whereas, when we further focus on higher charmonia such as $\psi(6S)$ and $\psi(7S)$, their theoretical mass will change about 100 MeV with the increase of a $\mu$ value from 0.11 to 0.15. This phenomenon can be understandable by the nature of the screened confining interaction.

\begin{figure}[h]
	\includegraphics[width=8cm,keepaspectratio]{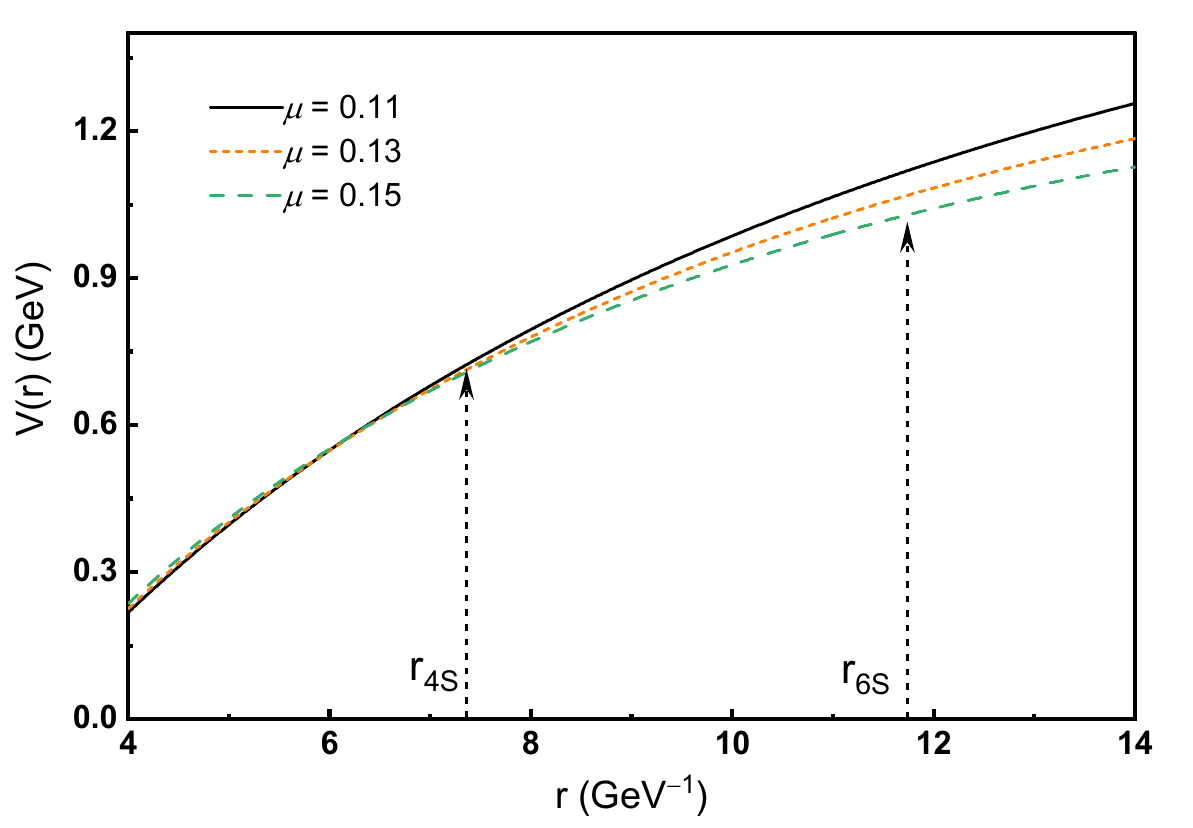}
	\caption{\label{fig:potential} The confinement plus one-gluon-exchange potential of a $c\bar{c}$ system with different $\mu$ values, where $\mu=0.11, 0.13, 0.15$ in units of GeV are presented. The RMS radius positions of $\psi(4S)$ and $\psi(6S)$ for $\mu=0.12$ are labeled as $r_{4S}$ and $r_{6S}$, respectively. }
\end{figure}

In Fig.~\ref{fig:potential}, we plot the screened confining plus one-gluon-exchange potential with different screened parameters, in which the root-mean-squared(RMS) radius of $\psi(4S)$ and $\psi(6S)$ are labeled as an illustration. One can see that the potential behavior is almost the same when the distance $r$ is smaller than the RMS radius of $\psi(4S)$, $r_{4S}$. However, the potentials with different $\mu$ values begin to obviously separate at the position of $r_{6S}$, at which the screened effect plays a very important role. As a result, the predictions of higher charmonia around 4.6 GeV based on the present experimental data in Table \ref{table:mu} have large uncertainties, and this also reflects on how poor we know about strong interactions. In short, a reasonable description of charmonium spectrum around 4.6 GeV is not an easy work, and we need more valuable experimental hints.

	\begin{table*}
  	\caption{ The charmonium mass spectrum with different $\mu$ values. Here, we take $\mu=0.11,0.12,0.13,0.14,0.15$ to show our results. The results in Ref. \cite{Wang:2019mhs} are also presented for comparison. All results are in units of MeV.}
  	\setlength{\tabcolsep}{4.5mm}{
  	\begin{tabular}{cccccccc}
			\bottomrule[1.2pt]
			\bottomrule[0.5pt]
			\textrm{} & \textrm{$\mu$=0.11} & \textrm{$\mu$=0.12} & \textrm{$\mu$=0.13} & \textrm{$\mu$=0.14} & \textrm{$\mu$=0.15} & Ref. \cite{Wang:2019mhs} & Expt. \cite{Tanabashi:2018oca}\\
			\hline
			\textrm{$\eta_c(1^1S_0)$} & 2984  & 2984 & 2984 & 2984 & 2984 & 2981 & 2983.9 $\pm$ 0.5 \\
			\textrm{$\psi(1^3S_1)$} & 3096 & 3096 & 3096 & 3097 & 3098 & 3096 & 3096.9 $\pm$ 0.5\\
			\textrm{$\chi_{c0}(1^3P_0)$} & 3449 & 3452 & 3455 & 3457 & 3462 & 3464 & 3414.71 $\pm$ 0.30 \\
			\textrm{$\chi_{c1}(1^3P_1)$} & 3515 & 3517 & 3520 & 3523 & 3528 & 3530 & 3510.67 $\pm$ 0.05 \\
			\textrm{$h_{c1}(1^1P_1)$} & 3523 & 3526 & 3528 & 3531 & 3536 & 3538 & 3525.38 $\pm$ 0.11 \\
			\textrm{$\chi_{c2}(1^3P_2)$} & 3555 & 3557 & 3560 & 3563 & 3568 & 3571 & 3556.17 $\pm$ 0.07 \\
			\textrm{$\eta_{c}(2^1S_0)$} & 3626 & 3669 & 3631 & 3634 & 3638 & 3642 & 3637.6 $\pm$ 1.2 \\
			\textrm{$\psi(2^3S_1)$} & 3667 & 3669 & 3672 & 3674 & 3679 & 3683 & 3686.097 $\pm$ 0.006\\
			\textrm{$\psi(1^3D_1)$} & 3808 & 3811 & 3818 & 3818 & 3824 & 3830 & 3778.1 $\pm$ 1.2\\
			\textrm{$\psi_{2}(1^3D_2)$} & 3827 & 3830 & 3833 & 3836 & 3842 & 3848 & 3822.2 $\pm$ 1.2 \\
			\textrm{$\psi_{3}(1^3D_3)$} & 3838 & 3841 & 3844 & 3847 & 3853 & 3859 & 3842.71 $\pm$ 0.16 $\pm$ 0.12 \cite{Aaij:2019evc} \\
			\textrm{$\chi_{c2}(2^3P_2)$} & 3937 & 3938 & 3939 & 3940 & 3944 & 3952 & 3927.2 $\pm$ 2.6 \\
			\textrm{$\psi(3^3S_1)$} & 4026 & 4025 & 4025 & 4024 & 4027 & 4035 & 4039 $\pm$ 1\\
			\textrm{$\psi(2^3D_1)$} & 4115 & 4114 & 4113 & 4112 & 4115 & 4125 & 4159 $\pm$ 20\\
			\textrm{$\psi(4^3S_1)$} & 4286 & 4279 & 4272 & 4264 & 4262 & 4274 & 4230 $\pm$ 8\\
			\textrm{$\psi(3^3D_1)$} & 4348 & 4340 & 4333 & 4324 & 4321 & 4334 & $\dotsb$ \\
			\textrm{$\psi(5^3S_1)$} & 4484 & 4470 & 4454 & 4437 & 4428 & 4443 & 4421 $\pm$ 4\\
			\textrm{$\psi(4^3D_1)$} & 4530 & 4514 & 4497 & 4479 & 4468 & 4484 & $\dotsb$ \\
			\textrm{$\psi(6^3S_1)$} & 4640 & 4615 & 4589 & 4562 & 4542 & $\dotsb$ & $\dotsb$ \\
			\textrm{$\psi(5^3D_1)$} & 4674 & 4648 & 4620 & 4591 & 4570 & $\dotsb$ & $\dotsb$ \\
			\textrm{$\psi(7^3S_1)$} & 4762 & 4726 & 4688 & 4649 & 4618 & $\dotsb$ & $\dotsb$\\
			\textrm{$\psi(6^3D_1)$} & 4788 & 4750 & 4711 & 4669 & 4636 & $\dotsb$ & $\dotsb$\\
			\textrm{$\chi^2/n$} & 30.1 & 25.2 & 21.8 & 24.4 & 22.0 & $\dotsb$ & $\dotsb$ \\
			\bottomrule[1.2pt]
		\end{tabular}\label{table:mu}}
  \end{table*}

\section{\label{section:fit}
Implication of charmoniumlike $Y$ states around 4.6 GeV
}

In the present experimental status, the charmoniumlike state $Y(4630)$ or $Y(4660)$ can be considered as a higher scaling point of a charmonium spectrum. Therefore, in this section, we will explore whether these $Y$ states can be treated as higher charmonia. We notice that both $Y(4220)$ and $Y(4660)$ have been discovered in the hidden charm process $e^+e^-\to\psi(2S)\pi^+\pi^-$ \cite{Wang:2007ea,Ablikim:2017oaf}. And if the nature of $Y(4220)$ is a charmonium $\psi(4S)$, then the $Y(4660)$ should also be a good candidate of a higher charmonium. Actually, there have been some theoretical works discussing the possibilities of $Y(4660)$ as a charmonium state after its experimental discovery \cite{Ding:2007rg,Li:2009zu}. In Ref. \cite{Li:2009zu}, Li and Chao suggested that $Y(4660)$ is the candidate of $6^3S_1$ $c\bar{c}$ state by analyzing their calculated mass, whose value is 4608 MeV and smaller than the experimental mass of $4652\pm10\pm8$ MeV \cite{Wang:2014hta}. From our results listed in Table \ref{table:mu}, the mass of $\psi(6S)$ is predicted to be 4542 to 4640 MeV with different $\mu$ values, which are also lower than the measured mass of $Y(4660)$. We also notice that the experimental mass difference $M_{Y(4660)}-M_{\psi(4415)}=231$ MeV is greater than $M_{\psi(4415)}-M_{Y(4220)}=191$ MeV. If the property of $Y(4660)$ is dominated by a $\psi(6S)$ state, according to experiences in the quark model, the above mass diferences are very abnormal because the mass gap should be expected to decrease with the increase of radial quantum numbers. Hence the real position of $\psi(6S)$ in charmonium spectrum is still an open problem.

When we continue to move our focus to another charmoniumlike structure $Y(4630)$, some interesting clues emerge. As early as 2008, the Belle has reported the observation of the $Y(4630)$ in the $e^+e^-\rightarrow\Lambda_c\bar{\Lambda}_c$ \cite{Pakhlova:2008vn}, whose mass and width are measured to be $4634^{+8+5}_{-7-8}$ and $92^{+40+10}_{-24-21}$ MeV, respectively. The production threshold of the $\Lambda_c\bar{\Lambda}_c$ mode is close to charmonia around 4.6 GeV focused by us, and hence, this mode is hopeful to provide some characteristic information. In addition to cross sections near the $Y(4630)$, the precise measurements of the cross section of the $e^+e^- \to \Lambda_c\bar{\Lambda}_c$ near threshold were recently released by BESIII \cite{Ablikim:2017lct}. Because of the limit of collision energy, the only four energy points of $\sqrt{s}$=4.5745, 4.580, 4.590, and 4.5995 GeV near $\Lambda_c\bar{\Lambda}_c$ threshold are measured, where the non-zero cross section is cleared. It is interesting that these precision data from BESIII show 
 a ``platform" structure near threshold, which is very different from that of Belle \cite{Pakhlova:2008vn}. At present, this phenomenon is not answered by any one of theoretical explanations. In Ref. \cite{Dai:2017fwx} , the authors considered final state interaction mechanism from $\Lambda_c\bar{\Lambda}_c$ and the contribution of a resonance state to revisit the present experimental data of $e^+e^-\rightarrow\Lambda_c\bar{\Lambda}_c$, where the data of Belle are well fitted. However, the latest data from BESIII cannot describe it yet. Therefore, this fact inspires us to propose a new perspective that the novel behavior near the threshold of the reaction $e^+e^-\rightarrow\Lambda_c\bar{\Lambda}_c$ may imply the existence of a potential resonance. The possible structure may be helpful to respond to the problem of the scaling point of a charmonium $\psi(6S)$ state. In order to verify our idea, we will reanalyze the related experimental data in the next subsection.


\subsection{Hint from $e^+e^-\to \Lambda_c\bar{\Lambda}_c$ and $e^+e^-\to \psi(2S)\pi^+\pi^-$ reactions}

As mentioned above, we suspect that there could be another charmoniumlike state around 4.6 GeV besides the known $Y(4630)$ in the reaction of $e^+e^-\rightarrow\Lambda_c\bar{\Lambda}_c$. Here, we assume that a line shape of the hidden charm process $e^+e^-\to \psi(2S)\pi^+\pi^-$ near the peak of $Y(4660)$ is also generated by two resonances, which has the same structure as that of the $e^+e^-\rightarrow\Lambda_c\bar{\Lambda}_c$. Based on this, we can make a combined fit to the experimental data of $e^+e^-\rightarrow\psi(2S)\pi^+\pi^-$ and $e^+e^-\rightarrow\Lambda_c\bar{\Lambda}_c$ measured by both Belle and BESIII.

The contributions of a genuine resonance to the $e^+e^-\rightarrow\psi(2S)\pi^+\pi^-$ and $e^+e^-\rightarrow\Lambda_c\bar{\Lambda}_c$ can be described by a phase space corrected Breit-Wigner distribution
\begin{eqnarray}
\mathcal{M}_{R}(Y)&=&\frac{\sqrt{12\pi\Gamma^{e^+e^-}_{Y}\times \mathcal{B}(Y\to\psi(2S)\pi^+\pi^-)\Gamma_{Y}}}{s-m_{Y}^2+im_{Y}\Gamma_{Y}} \nonumber \\
&&\times \sqrt{\frac{\Phi_{2\to3}(s)}{\Phi_{2\to3}(m_{Y}^2)}}
\end{eqnarray}
and
\begin{eqnarray}
\mathcal{M}_{R}^{\prime}(Y)=\frac{\sqrt{12\pi\Gamma^{e^+e^-}_{Y}\times \mathcal{B}(Y\to\Lambda_c\bar{\Lambda}_c)\Gamma_{Y}}}{s-m_{Y}^2+im_{Y}\Gamma_{Y}}
\sqrt{\frac{\Phi_{2\to2}(s)}{\Phi_{2\to2}(m_{Y}^2)}},
\end{eqnarray}
respectively, where $Y$ denotes the intermediate resonance, and $\Phi_{2\to3}(s)$ and $\Phi_{2\to2}(s)$ are the phase spaces of the process $e^+e^-\rightarrow\psi(2S)\pi^+\pi^-$ and $e^+e^-\rightarrow\Lambda_c\bar{\Lambda}_c$, respectively. Here, we define a free parameter $\mathcal{R}_{Y}$, which is equal to the product of di-lepton width $\Gamma^{e^+e^-}_{Y}$ and  {branching ratio $\mathcal{B}(Y \to final~states)$. } It is worth noting that the cross section of $e^+e^-\rightarrow\Lambda_c\bar{\Lambda}_c$ for extremely close to the production threshold is non-zero. Hence, several non-resonant contributions are necessarily considered. For simplicity, we adopt a parameterized threshold function to describe a scattering amplitude of all non-resonant terms \cite{Pakhlova:2008vn}, which is
\begin{equation}
\mathcal{M}_{NoR}=g_{NoR}(\sqrt{s}-2m_{\Lambda_c})^{\frac{1}{2}}e^{-(a\sqrt{s}+bs)},
\end{equation}
where $g_{NoR}$, $a$, and $b$ are free parameters, and $\sqrt{s}$ is the energy of center of mass. The total amplitude of the $e^+e^-\rightarrow\psi(2S)\pi^+\pi^-$ and $e^+e^-\rightarrow\Lambda_c\bar{\Lambda}_c$ can be written as
\begin{eqnarray}
&&\mathcal{M}_{\psi(2S)\pi^+\pi^-}^{\mathrm{Total}}=\mathcal{M}_{R}(Y_1)+e^{i\theta}\mathcal{M}_{R}(Y_2), \\
&&\mathcal{M}_{\Lambda_c\bar{\Lambda}_c}^{\mathrm{Total}}=\mathcal{M}_{NoR}+e^{i\phi_1}\mathcal{M}^{\prime}_{R}(Y_1)+e^{i\phi_2}\mathcal{M}^{\prime}_{R}(Y_2),
\end{eqnarray}
where $\theta$, $\phi_1$, and $\phi_2$ are the phase angles among different amplitudes.

\begin{figure}
	\includegraphics[width=7cm,keepaspectratio]{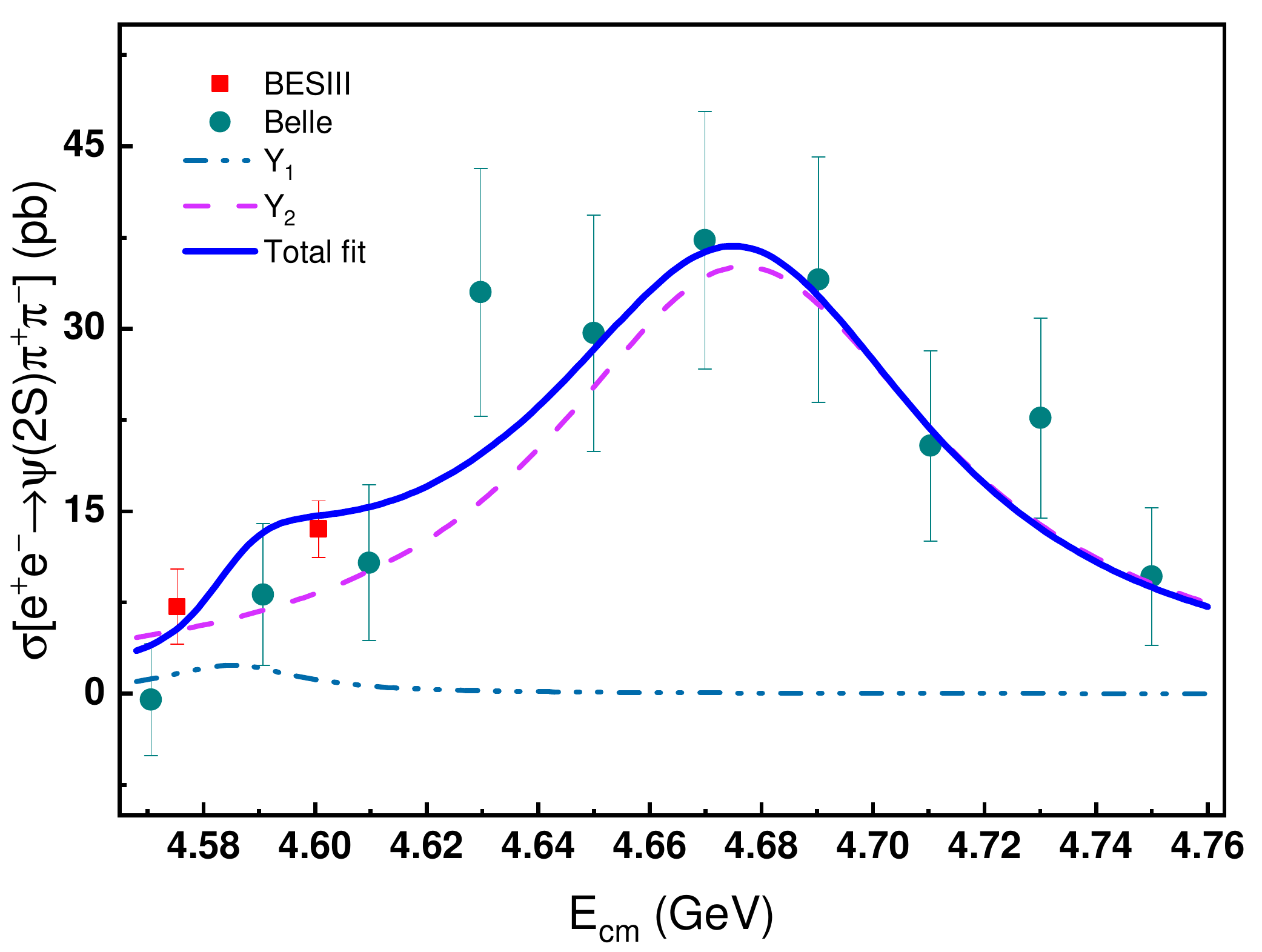}
	\caption{ Our fit to the experimental data of $e^+e^-\rightarrow \psi(2S)\pi^+\pi^-$} from Belle and BESIII \cite{Wang:2014hta,Ablikim:2017oaf}.
	\label{fig:2spipi}
\end{figure}

\begin{figure}
	\includegraphics[width=7cm,keepaspectratio]{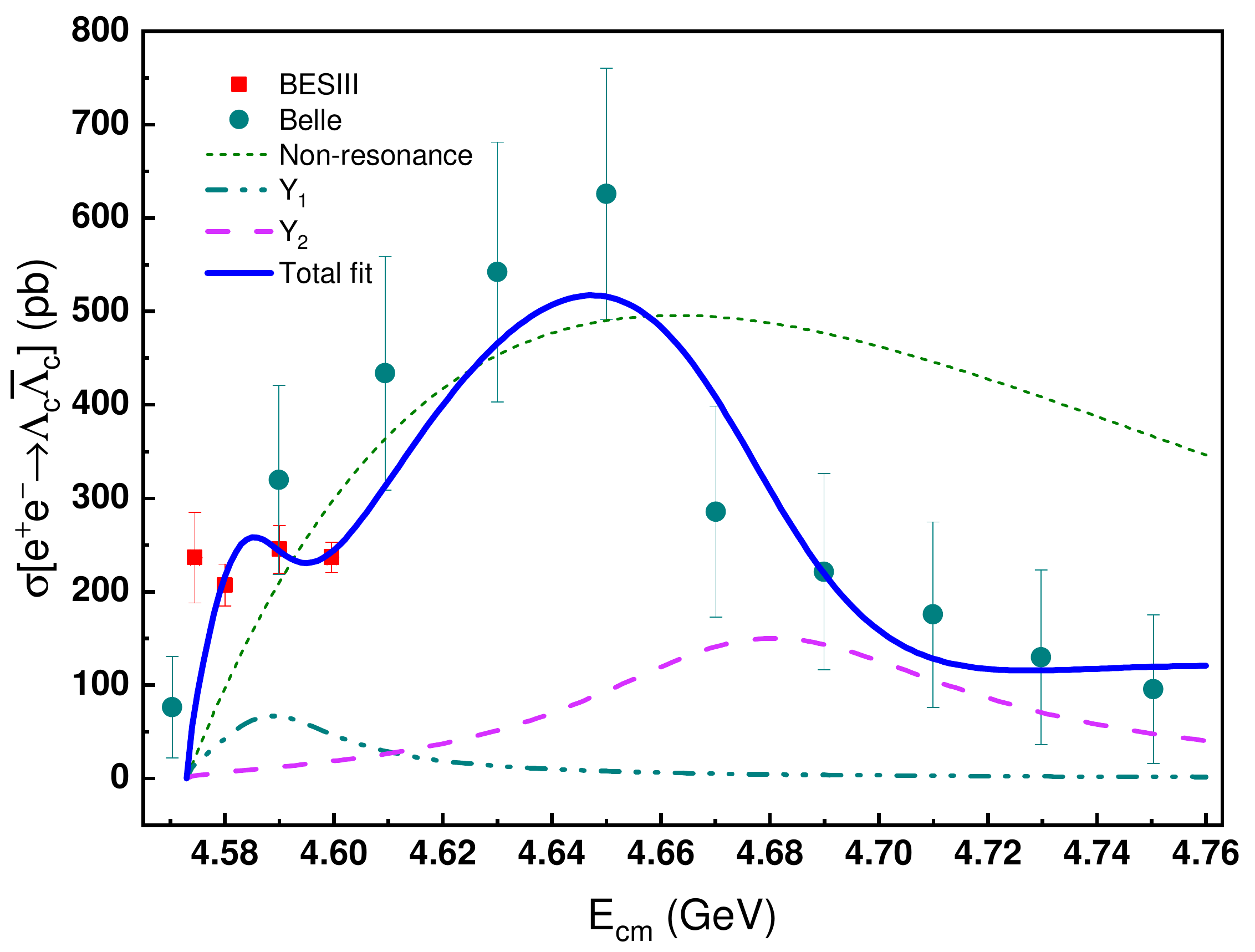}
	\caption{ Our fit to the experimental data of $e^+e^-\rightarrow \Lambda_c\bar{\Lambda}_c$ from Belle and BESIII \cite{Pakhlova:2008vn,Ablikim:2017lct}.}
	\label{fig:lambdac}
\end{figure}

With the above preparation, a combined fit to the experimental data both of the $e^+e^-\rightarrow\psi(2S)\pi^+\pi^-$ and $e^+e^-\rightarrow\Lambda_c\bar{\Lambda}_c$ \cite{Wang:2014hta,Ablikim:2017oaf,Pakhlova:2008vn,Ablikim:2017lct} is achieved, and the corresponding fitted line shapes are shown in Fig.~\ref{fig:2spipi} and Fig.~\ref{fig:lambdac}. The relevant fitted parameters are listed in Table \ref{table:fitparameter}, where the $\chi^2/d.o.f$=1.49 is obtained. To our surprise, the ``platform" behavior of experimental data of $e^+e^-\rightarrow\Lambda_c\bar{\Lambda}_c$ near threshold can be indeed reproduced in our proposed two-resonance scheme, which can be seen in Fig. \ref{fig:lambdac}. Meanwhile, the experimental data of $e^+e^-\rightarrow\psi(2S)\pi^+\pi^-$ are also described well as shown in Fig. \ref{fig:2spipi}, where a precise data point at 4.6 GeV from BESIII also provides the possible evidence to support our conjecture of the existence of a new charmoniumlike structure around 4.6 GeV. Anyway, more precise experimental measurements are very important for clarifying the structures 
in this energy region.


In our scheme, the resonance parameters of two charmoniumlike structures $Y_1$ and $Y_2$ are fitted to be
\begin{eqnarray}
    m_{Y_1}=4585\pm2\, \mathrm{MeV},     \quad\quad \Gamma_{Y_1}=29.8\pm8.0\, \mathrm{MeV}, \nonumber \\
    m_{Y_2}=4676\pm7\, \mathrm{MeV},     \quad\quad \Gamma_{Y_2}=85.7\pm15.0\, \mathrm{MeV},
\end{eqnarray}
where the mass of the former state is 4585 MeV, which is consistent with our estimate of 4542 to 4640 MeV for the mass of charmonium $\psi(6S)$. In addition, its mass is near 80 MeV smaller than that of $Y(4660)$, and the mass gap puzzle as mentioned above can be also understood if the lower $Y_1$ instead of $Y(4660)$ is the candidate for $\psi(6S)$. Thus, the underlying $Y_1$ state as $\psi(6S)$ can be treated as a higher scaling point to determine the screened parameter in our unquenched potential model. The latter state can be related to our known $Y(4660)$ and $Y(4630)$, which can be naturally assigned to a charmonium $\psi(5D)$ from a mass spectrum point of view. In the early studies \cite{Bugg:2008sk,Guo:2010tk,Cotugno:2009ys}, theorists have paid a lot of attention to discuss whether the $Y(4630)$ and $Y(4660)$ observed in different decay modes have the same structure. In our analysis, we support that these two charmoniumlike $Y$ states are probably identified as the same resonance $Y_2$.

\begin{table*}[t]
\caption{ The fitted parameters in a combined fit to experimental data of the $e^+e^-\rightarrow\psi(2S)\pi^+\pi^-$ and $e^+e^-\rightarrow\Lambda_c\bar{\Lambda}_c$ \cite{Wang:2014hta,Ablikim:2017oaf,Pakhlova:2008vn,Ablikim:2017lct}, where the $\chi^2/d.o.f$=1.49 is obtained.}
\begin{ruledtabular}
\begin{tabular}{ccccccccccccccc}
  Parameter   & $g_{NoR}$ & $a$ & $b$ & $\mathcal{R}^{Y_1}_{\Lambda_c\bar{\Lambda}_c}$& $\phi_1$& $\mathcal{R}^{Y_2}_{\Lambda_c\bar{\Lambda}_c}$ & $\phi_2$& $\mathcal{R}^{Y_1}_{\psi(2S)\pi^+\pi^-}$ &$\mathcal{R}^{Y_2}_{\psi(2S)\pi^+\pi^-}$& $\theta$& $m_{Y_1}$& $\Gamma_{Y_1}$ &$m_{Y_2}$& $\Gamma_{Y_2}$\\
  (Unit)   & ($\mathrm{GeV}^{-\frac{3}{2}}$) &($\mathrm{GeV}^{-1}$)&($\mathrm{GeV}^{-2}$)&(eV)&(rad)&(eV)&(rad)&(eV)&(eV)&(rad)&(GeV)&(GeV)&(GeV)&(GeV)\\
     \hline
  Value & $15.9\times10^3$ & $0.970$ & $0.494$& $2.66$& 3.39& 19.0& 3.56& 0.100& 4.48 & 2.42& 4.585& 0.0299& 4.676& 0.0857 \\
  Error($\pm$) &$0.5\times10^3$&0.008&0.002&1.27&0.14&4.6&0.29&0.032&0.69&0.50&0.002&0.0080&0.007&0.0150
\end{tabular}
\end{ruledtabular}
\label{table:fitparameter}
\end{table*}

\subsection{$6S$-$5D$ mixing scheme}

\begin{figure}[t]
    \includegraphics[width=8cm,keepaspectratio]{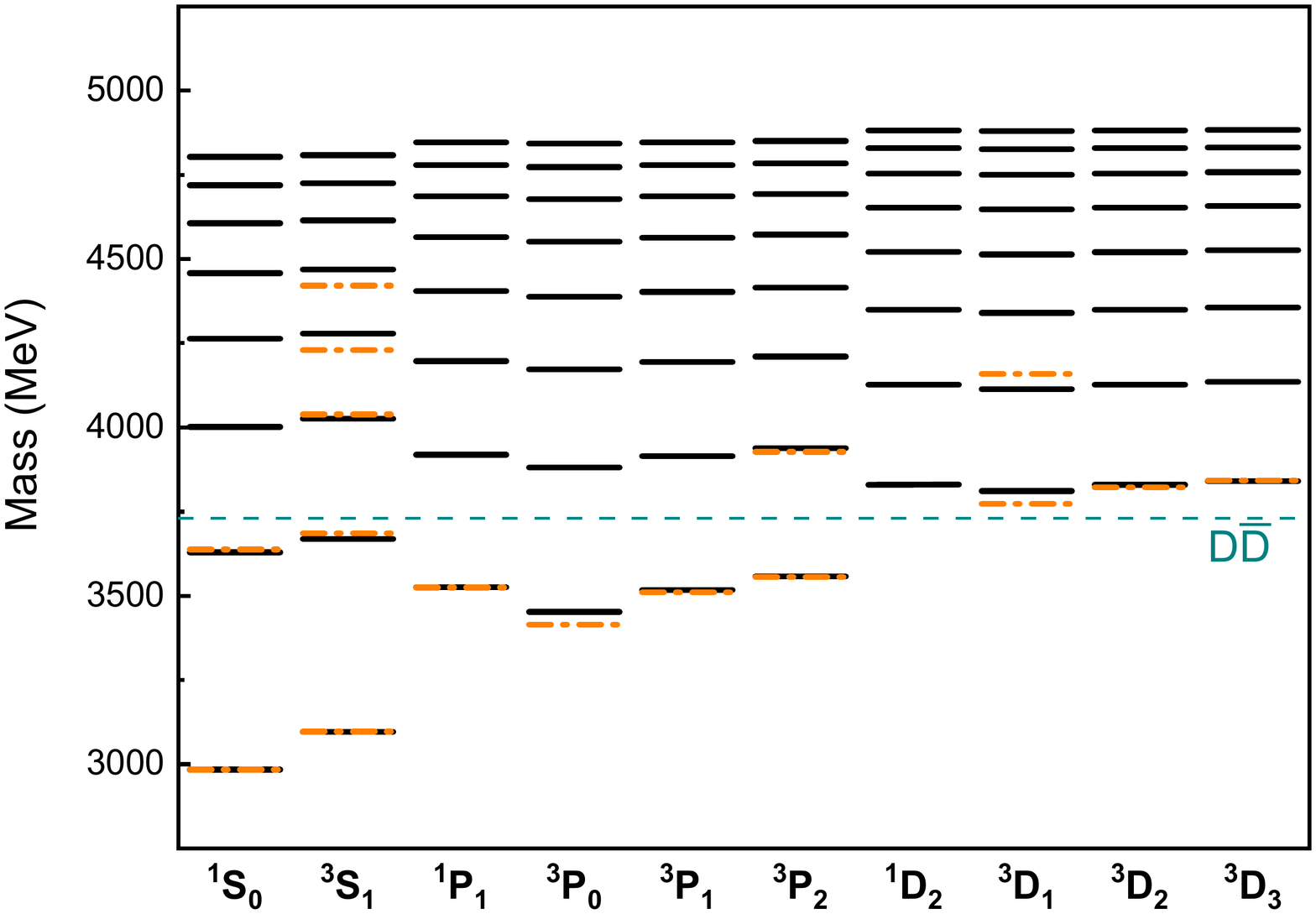}
    \caption{ The charmonium spectrum, where the screening parameter is taken as $\mu$=0.12. The orange dash-dotted and black solid lines correspond to experimental results and model predictions, respectively. }
    \label{fig:spectrum}
    \end{figure}

In the above subsection, we carefully investigate the possibility whether the fitted charmoniumlike resonances $Y_1$ and $Y_2$ can be treated as two missing charmonium $\psi(6S)$ and $\psi(5D)$, respectively. In our previous studies \cite{Wang:2019mhs}, we found that the contribution of $4S$-$3D$ mixture is very significant for explaining both $Y(4220)$ and predicted $\psi(4380)$. Hence, we further study the properties of charmonium $\psi(6S)$ and $\psi(5D)$ in the $6S$-$5D$ mixture scheme \footnote{ {As indicated in Ref. \cite{Rosner:2001nm}, the possible origins of $S$-$D$ wave mixture of hadron states include the tensor interaction and coupled channel effects. In the charmonium family, a typical example of the $S$-$D$ mixture is $\psi(3770)$, which is generally assigned to be as a $1^3D_1$ charmonium state. However, many experimental evidences show that $\psi(3770)$ should be a mixing charmonium with the component of $1^3D_1$ and $2^3S_1$ state, where the mixing angle is suggested to be $(12\pm2)^{\circ}$ in Ref. \cite{Rosner:2001nm}. For higher charmonia around 4.6 GeV, we have to adopt a phenomenological way to study their $6S$-$5D$ mixture scheme due to the lack of experimental information. Of course, the research of mixing phenomena in charmonium system is very interesting and can be considered in the future.} }.
Under the $S$-$D$ mixing framework, we introduce
\begin{equation}
	\begin{pmatrix}
	|\psi'_{6S-5D}\rangle\\
	|\psi''_{6S-5D}\rangle
	\end{pmatrix}
	=
	\begin{pmatrix}
	\mathrm{cos}\theta  & \mathrm{sin}\theta\\
	-\mathrm{sin}\theta & \mathrm{cos}\theta
	\end{pmatrix}
	\begin{pmatrix}
	|6^3S_1\rangle\\
	|5^3D_1\rangle
	\end{pmatrix}.
\end{equation}
Here, $\theta$ denotes the mixing angle, and the mass eigenvalues of $\psi'_{6S-5D}$ and $\psi''_{6S-5D}$ are determined by the masses of two basis vectors $m_{6S}$, $m_{5D}$, and the mixing angle $\theta$, i.e.,
\begin{equation}
	m^2_{\psi'_{6S-5D}}=\frac{1}{2}\left(m^2_{6S}+m^2_{5D}-\sqrt{(m^2_{5D}-m^2_{6S})^2\mathrm{sec}^22\theta}\right),
\end{equation}
\begin{equation}
	m^2_{\psi''_{6S-5D}}=\frac{1}{2}\left(m^2_{6S}+m^2_{5D}+\sqrt{(m^2_{5D}-m^2_{6S})^2\mathrm{sec}^22\theta}\right),
\end{equation}
where the input of $m_{6S}$ and $m_{5D}$ are dependent on our prediction from the unquenched potential model. We note that when taking screened parameter $\mu=0.12$, the masses of pure $6S$ and $5D$ $c\bar{c}$ states are 4615 and 4648 MeV as shown in Table \ref{table:mu}, respectively. This result roughly coincides with the mass of two fitted resonances, so we choose the screened parameter $\mu=0.12$ to describe higher charmonia around 4.6 GeV. The complete parameters adopted in the unquenched potential model are listed as follows,
\begin{eqnarray}
\epsilon_c&=&-0.084,    \quad\quad \epsilon_t=0.012,\nonumber\\
\epsilon_{sov}&=&-0.053,\quad\quad \epsilon_{sos}=0.083,\nonumber\\
b&=&0.238\, \mathrm{GeV}^2,              \quad\quad c=-0.337\, \mathrm{GeV},\nonumber\\
m_c&=&1.65\,\mathrm{GeV},\quad\quad \mu=0.12\,\mathrm{GeV}.\nonumber
\end{eqnarray}
The charmonium spectrum with this group of parameters is also depicted in Fig.~\ref{fig:spectrum} for reference.

\begin{figure}[htbp]
	\includegraphics[width=7cm,keepaspectratio]{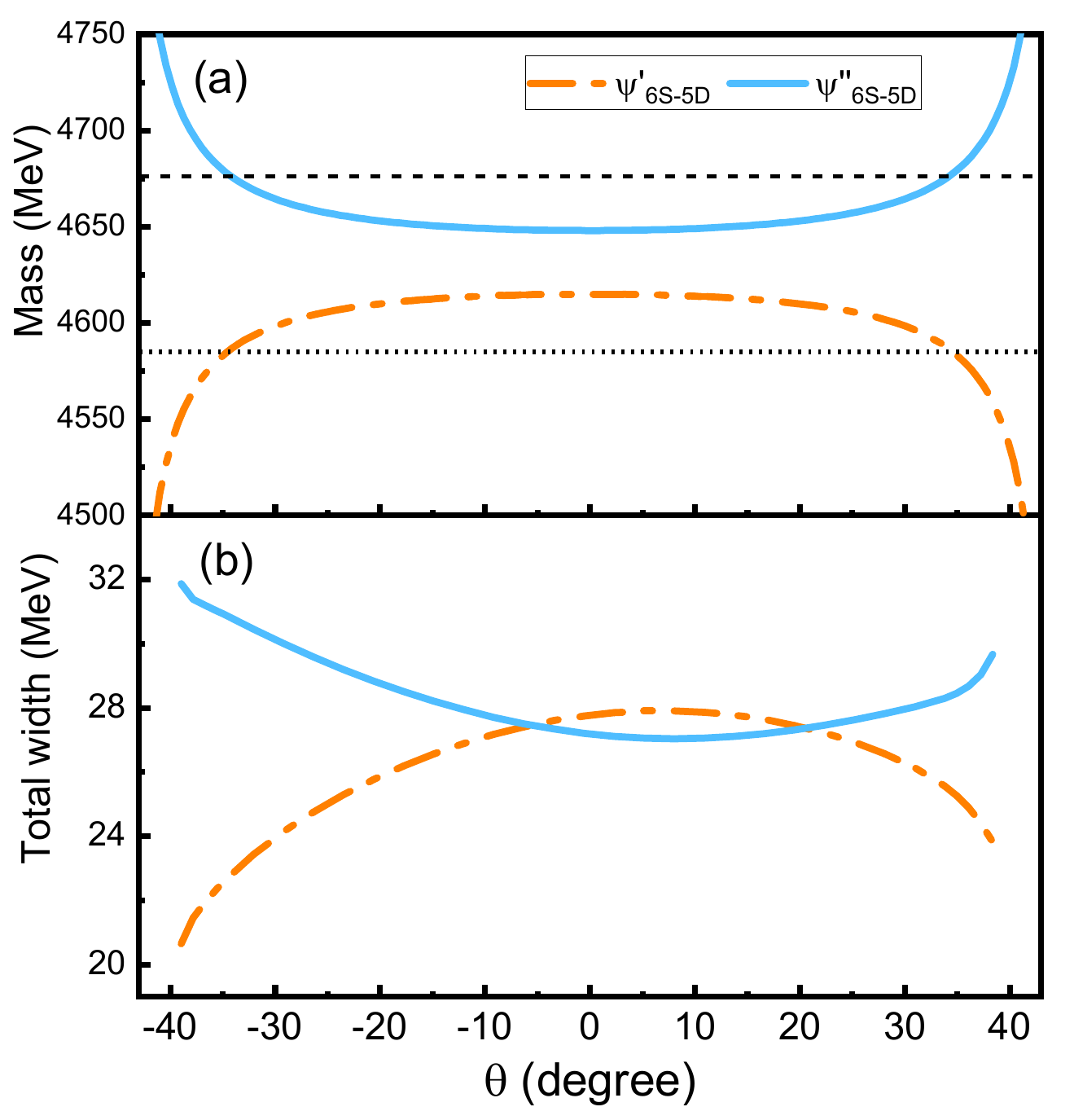}
	\caption{The dependence of masses and widths of $\psi^{\prime}_{6S-5D}$ and $\psi^{\prime\prime}_{6S-5D}$ on mixing angel $\theta$ in the $6S$-$5D$ mixing scheme. The dotted and dashed lines correspond to the masses of fitted resonances $Y_1$ and $Y_2$ in Section ~\ref{section:fit}, respectively. }
	\label{fig:mixing}
\end{figure}

In Fig.~\ref{fig:mixing}, we present the dependence of masses of $\psi'_{6S-5D}$ and $\psi''_{6S-5D}$ on mixing angle $\theta$. As the $\theta$ increases gradually, it can be seen that the mass of $\psi'_{6S-5D}$ is decreased and the mass of $\psi''_{6S-5D}$ is oppositely enhanced. When mixing angle $\theta=\pm34^{\circ}$, we were surprised to find that the theoretical masses of $\psi'_{6S-5D}$ and $\psi''_{6S-5D}$ are perfectly consistent with our fitting resonance masses, whose comparisons are shown in Table~\ref{table:mixing}. It is no doubt that this mass consistency provides us great confidence to believe these charmoniumlike $Y$ states around 4.6 GeV can be treated as higher members in $J/\psi$ family.

To further identify the nature of two $Y$ structures around 4.6 GeV as charmonia, we also study the open-charm strong decay behaviors of $\psi'_{6S-5D}$ and $\psi''_{6S-5D}$, where the quark pair creation (QPC) model \cite{Micu:1968mk,LeYaouanc:1977gm} is adopted. The QPC model is a very successful phenomenological model in treating the OZI-allowed two-body strong decays of hadrons, and we first concisely illustrate the model.

In the QPC model, transition matrix of the process $A \rightarrow B+C$ is written as $\langle BC|{\cal T}|A\rangle=\delta^3(\boldsymbol{P}_B+\boldsymbol{P}_C){\cal M}^{M_{J_A}M_{J_B}M_{J_C}}(\boldsymbol{P})$, where the transition operator ${\cal T}$ describes a quark-antiquark pair creation from the vacuum and has the form
\begin{equation}
\begin{split}
{\cal T}&=-3\gamma\sum_{m,i,j}^{}\langle 1m;1-m|00\rangle\int
d\boldsymbol{p}_3d\boldsymbol{p}_4\delta^3(\boldsymbol{p}_3+\boldsymbol{p}_4)\\
&\quad\times {\cal Y}_{1m}\left(\frac{\boldsymbol{p}_3-\boldsymbol{p}_4}{2}\right)\chi^{34}_{1,-m}\phi^{34}_{0}(\omega^{34}_0)_{ij}b^{\dagger}_{3i}(\boldsymbol{p}_3)b^{\dagger}_{4j}(\boldsymbol{p}_4).
\end{split}
\end{equation}
A dimensionless constant $\gamma$ is introduced to describe the strength of creating the quark pair $u\bar{u}$ or $d\bar{d}$ from the vacuum, which can be fixed by experimental data. The $\chi^{34}_{1,-m}$ is a spin-triplet state, and $\omega^{34}_{0}$ and $\phi^{34}_{0}$ denote $SU(3)$ color and flavor singlets, respectively. ${\cal Y}_{lm}({\boldsymbol{p}})\equiv |\boldsymbol{p}|^lY_{lm}(\theta_p,\phi_p)$ denotes the $l$-th solid harmonic polynomial. The helicity amplitude ${\cal M}^{M_{J_A}M_{J_B}M_{J_C}}(\boldsymbol{P})$ could be related to the partial wave amplitude by the Jacob-Wick formula \cite{Jacob:1959at},
\begin{equation}
\begin{split}
{\cal M}^{JL}(A\rightarrow BC)& =\frac{\sqrt{(2L+1)}}{2J_A+1}\sum_{M_{J_B}M_{J_C}}
\langle L0;JM_{J_A}|J_AM_{J_A}\rangle\\
& \quad \times \langle J_BM_{J_B};J_CM_{J_C}|J_AM_{J_A}\rangle\\
& \quad \times {\cal M}^{M_{J_A}M_{J_B}M_{J_C}}(\boldsymbol{P}),
\end{split}
\end{equation}
where $\boldsymbol{L}$ is the orbital angular momentum between final states $B$ and $C$, and $\boldsymbol{J}=\boldsymbol{J}_B+\boldsymbol{J}_C$. The general partial width of the $A \rightarrow BC$ reads as
\begin{equation}
\Gamma_{A\rightarrow BC}=\pi^2\frac{|\boldsymbol{P}_B|}{m^2_A}\sum_{J,L}|{\cal M}^{JL}(\boldsymbol{P})|^2.
\end{equation}

In our calculation, the radial wave function of an initial charmonium state is directly obtained from the unquenched potential model. The final charmed or charmed-strange meson wave functions and the corresponding masses are taken from Refs. \cite{Song:2015nia,Song:2015fha}. The constituent quark masses $m_c$, $m_u=m_d$, and $m_s$ are taken as 1.65, 0.22, and 0.419 GeV, respectively. The dimensionless parameter $\gamma=5.84$ is the same as in Ref. \cite{Wang:2019mhs}, and the strength for creating $s\bar{s}$ from the vacuum satisfies the relation of $\gamma_s=\gamma/\sqrt{3}$ \cite{LeYaouanc:1977gm}.

\begin{table}[h]
\caption{A comparison between theoretical masses and widths in the $6S$-$5D$ mixing scheme, and the resonance parameters derived from a combined fit to the $e^+e^-\rightarrow\psi(2S)\pi^+\pi^-$ and $e^+e^-\rightarrow\Lambda_c\bar{\Lambda}_c$. Here, the mixing angle is taken to be $\pm34^{\circ}$.}
\begin{ruledtabular}
\begin{tabular}{ccccc}
     & \multicolumn{2}{c}{$\theta=-34^{\circ}(34^{\circ})$} & \multicolumn{2}{c}{Fit}\\
     & $M  (\textrm{MeV})$ & $\Gamma  (\textrm{MeV})$ & $M  (\textrm{MeV})$ & $\Gamma  (\textrm{MeV})$\\
     \hline
    $\psi'_{6S-5D}$ & 4587(4587) & 23(25) & 4585$\pm$2 & 29.8$\pm$8\\
    $\psi''_{6S-5D}$ & 4675(4675) & 31(28) & 4676$\pm$7 & 85.7$\pm$15
\end{tabular}
\end{ruledtabular}
\label{table:mixing}
\end{table}

We firstly calculate the total and partial widths of the open charm two-body strong decays of pure $\psi(6S)$ and $\psi(5D)$ states, which are summarized in Table \ref{tab:table1}. The total width of $\psi(6S)$ is estimated to be 28.5 MeV, which is in full accordance with the fitted width of $29.8\pm8$ MeV of the charmoniumlike $Y_1$ state. After considering the $S$-$D$ mixing, we find the total widths of $\psi'_{6S-5D}$ and $\psi''_{6S-5D}$ are not sensitive to the mixing angle as shown in Fig. \ref{fig:mixing}. Thus, both mass and width of the $\psi'_{6S-5D}$ dominated by the $c\bar{c}$ component of $6^3S_1$ are consistent with the $Y_1$ state. This gives us
a very strong support to our conjecture that there exists a charmonium candidate with a lower mass beside the $Y(4660)$ in the $e^+e^-\to \psi(2S)\pi^+\pi^-$ and $Y(4630)$ in the $e^+e^- \to \Lambda_c \bar{\Lambda}_c$.

When the mixing angle of the $6S$-$5D$ mixture is $\pm34^{\circ}$, the mass and total width of $\psi''_{6S-5D}$ are 4675 and 30 MeV, respectively, where the theoretical mass is close to 4676$\pm$7 MeV of the fitted $Y_2$ state, but the width is smaller than the fitted result of $85.7\pm15$ MeV. However, we should emphasize here that the width of $Y_2$ resonance in our present fit may be not accurate due to the possible influences from higher charmonia above the peak of $\psi''_{6S-5D}$. For example, we notice that the mass of pure $\psi(7S)$ state is 4.726 GeV as listed in Table \ref{tab:table1}, which can easily interfere with the resonance $Y_2$ to change its Breit-Wigner distribution. The present experimental data cannot allow us to include their contributions in the fit of the $e^+e^-\rightarrow\psi(2S)\pi^+\pi^-$ and $e^+e^-\rightarrow\Lambda_c\bar{\Lambda}_c$. Hence, the richer experimental measurements will be helpful to clarify this point in the future.

{Finally, we briefly discuss two observed decay channels $\psi'_{6S-5D}$/$\psi''_{6S-5D} \to \psi(2S)\pi^+\pi^-$ and $\psi'_{6S-5D}$/$\psi''_{6S-5D} \to \Lambda_c\bar{\Lambda}_c$. In our combined fit to experimental data of the $e^+e^-\rightarrow\psi(2S)\pi^+\pi^-$ and $e^+e^-\rightarrow\Lambda_c\bar{\Lambda}_c$, we obtain four parameters $\mathcal{R}^{Y_1}_{\psi(2S)\pi^+\pi^-}$, $\mathcal{R}^{Y_2}_{\psi(2S)\pi^+\pi^-}$, $\mathcal{R}^{Y_1}_{\Lambda_c\bar{\Lambda}_c}$ and $\mathcal{R}^{Y_2}_{\Lambda_c\bar{\Lambda}_c}$, which correspond to the product of di-lepton width and branching ratio of decay mode $\psi(2S)\pi^+\pi^-$ or $\Lambda_c\bar{\Lambda}_c$ of charmonium  $\psi_{6S-5D}$ state. Generally, the width of the di-lepton of charmonia has order of keV \cite{Tanabashi:2018oca}. Thus, we can roughly estimate the branching ratio of $\psi'_{6S-5D}$/$\psi''_{6S-5D} \to \psi(2S)\pi^+\pi^-$ and $\psi'_{6S-5D}$/$\psi''_{6S-5D} \to \Lambda_c\bar{\Lambda}_c$ to be around $10^{-4}\sim10^{-3}$ and $10^{-3}\sim10^{-2}$, respectively, according to the fitted value of $\mathcal{R}^{Y}$ listed in Table \ref{table:fitparameter}. These results are consistent with the charmonium decay behavior since the $\Lambda_c\bar{\Lambda}_c$ is an open charm decay mode while the $\psi(2S)\pi^+\pi^-$ is a hidden charm final states. In future work, we will try to explore the dynamics mechanisms involved in these decay channels, which is another approach to decode the properties of charmonia around 4.6 GeV. }

\subsection{The new resonance observed in the process $e^+e^-\to D_s^+ D_{s1}(2536)^-+c.c.$ by Belle}

Very recently, the Belle collaboration reported the observation of a new resonance in the process of $e^+e^-\to D_s^+ D_{s1}(2536)^-+c.c.$ via initial-state radiation \cite{Jia:2019gfe}. The observed resonance with a mass of $4625.9^{+6.2}_{-6.0}({\rm stat.})\pm0.4({\rm syst.})$ MeV and a width of $49.8^{+13.9}_{-11.5}({\rm stat.})\pm4.0({\rm syst.})$ MeV has a certain quantum number $J^{PC}=1^{--}$, which is consistent with charmonium $J/\psi$ family. This is the first observation of the charmoniumlike state around 4.6 GeV in open charm decay channel. We also notice that the resonance parameters of this new structure are close to those of earlier reported $Y(4630)$ and $Y(4660)$. Hence, it is necessary to explore whether the latest experimental results from Belle can be understood by our theoretical framework.

\begin{figure}[htbp]
	\includegraphics[width=7cm,keepaspectratio]{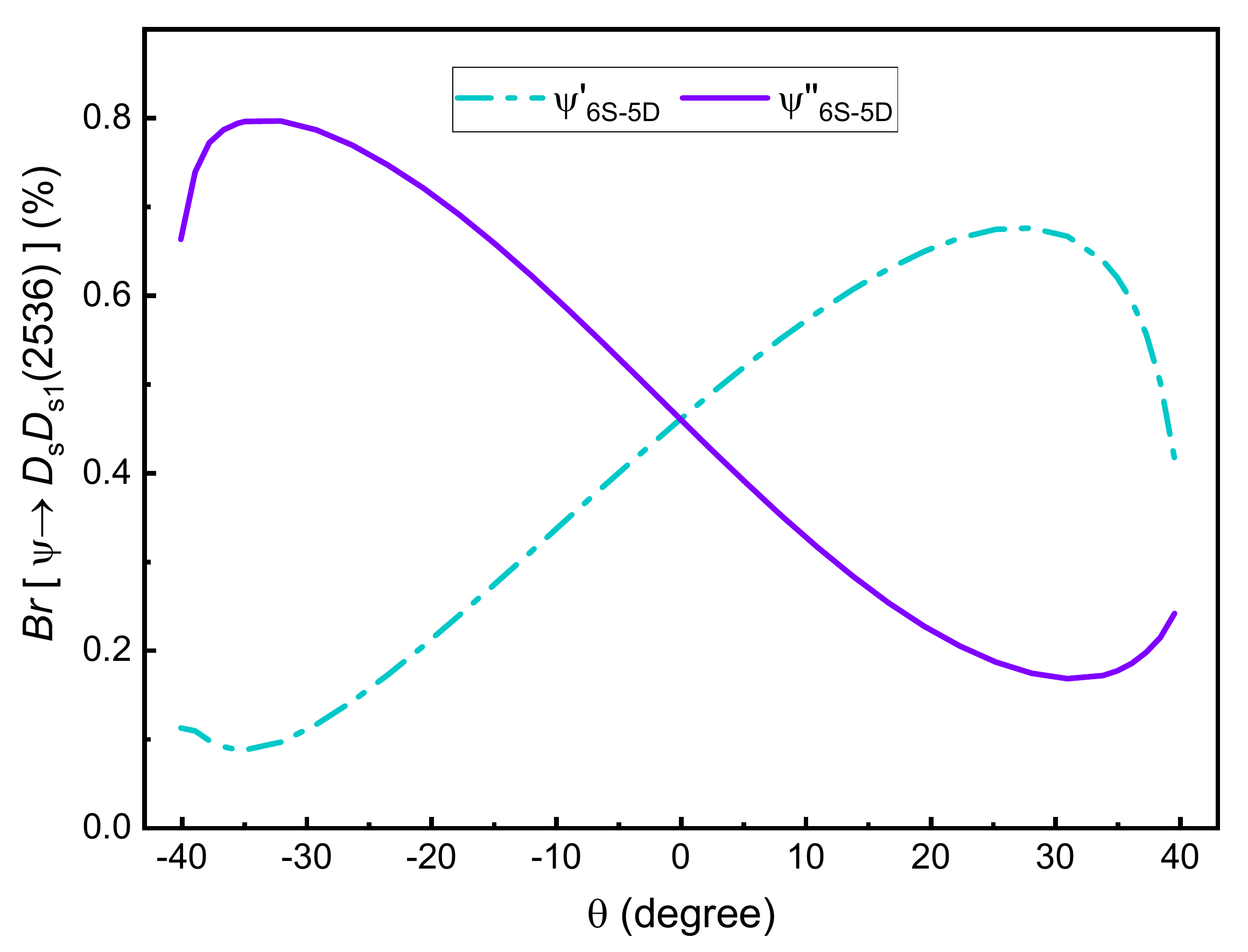}
	\caption{The dependence of a branching ratio of the $\psi^{\prime}_{6S-5D}/\psi^{\prime\prime}_{6S-5D}\to D_s D_{s1}(2536)$ on the mixing angle $\theta$ in the $6S$-$5D$ mixing scheme.}
	\label{fig:dsds1the}
	\end{figure}

We firstly calculate a branching ratio of the decay $\psi^{\prime}_{6S-5D}\to D_s D_{s1}(2536)$ and $\psi^{\prime\prime}_{6S-5D}\to D_s D_{s1}(2536)$ in the $6S$-$5D$ mixing scheme, whose dependence on the mixing angle $\theta$ is plotted in Fig. \ref{fig:dsds1the}. It can be seen that there are totally different decay behaviors of the $D_sD_{s1}(2536)$ mode in the negative and positive mixing angle. When mixing angle is taken as negative typical value $-34^{\circ}$, their branching ratios are
\begin{eqnarray}
\mathcal{B}(\psi^{\prime}_{6S-5D}\to D_s^+ D_{s1}^-(2536)+c.c.)=0.09\%, \nonumber \\
\mathcal{B}(\psi^{\prime\prime}_{6S-5D}\to D_s^+ D_{s1}^-(2536)+c.c.)=0.80\%. \label{eq-dsds11}
\end{eqnarray}
However, the positive mixing angle $+34^{\circ}$ corresponds to
\begin{eqnarray}
\mathcal{B}(\psi^{\prime}_{6S-5D}\to D_s^+ D_{s1}^-(2536)+c.c.)=0.64\%, \nonumber \\
\mathcal{B}(\psi^{\prime\prime}_{6S-5D}\to D_s^+ D_{s1}^-(2536)+c.c.)=0.17\%. \label{eq-dsds12}
\end{eqnarray}
We can see that the contribution of the $D_sD_{s1}(2536)$ mode in the decays of charmonium $\psi^{\prime}_{6S-5D}$ and $\psi^{\prime\prime}_{6S-5D}$ can reach the order of magnitude from $10^{-3}$ to $10^{-2}$, which is not contradictory with experimental observation of Belle. Although a resonance is discovered in the process $e^+e^-\to D_s D_{s1}(2536)$, we still adopt the two-resonance scheme to simply analyze the experimental data of Belle for the consistency from our theoretical view.
Because the production threshold of the $D_s D_{s1}(2536)$ mode is not far from the 4.6 GeV, both the non-resonant and resonance contributions will be considered in our analysis for cross sections of the $e^+e^-\to D_s D_{s1}(2536)$ reaction by Belle, which refers to the treatments in Refs. \cite{Chen:2015bft,Chen:2017uof}. The resonance contributions from $\psi^{\prime}_{6S-5D}$ and $\psi^{\prime\prime}_{6S-5D}$ can be fixed by the theoretical estimates, where two kinds of a mixing angle, $-34^{\circ}$ and $+34^{\circ}$, are considered. Their branching ratios of the $D_sD_{s1}(2536)$ mode in positive and negative angle have summarized in Eqs. (\ref{eq-dsds11})-(\ref{eq-dsds12}). In addition, since the di-lepton width of a charmonium is generally a keV order of magnitude \cite{Tanabashi:2018oca}, we can take $\Gamma(\psi^{\prime}_{6S-5D}/\psi^{\prime\prime}_{6S-5D}\to e^+e^-)$ to be 1 keV as a reference value.

    \begin{figure}[htbp]
	\includegraphics[width=8cm,keepaspectratio]{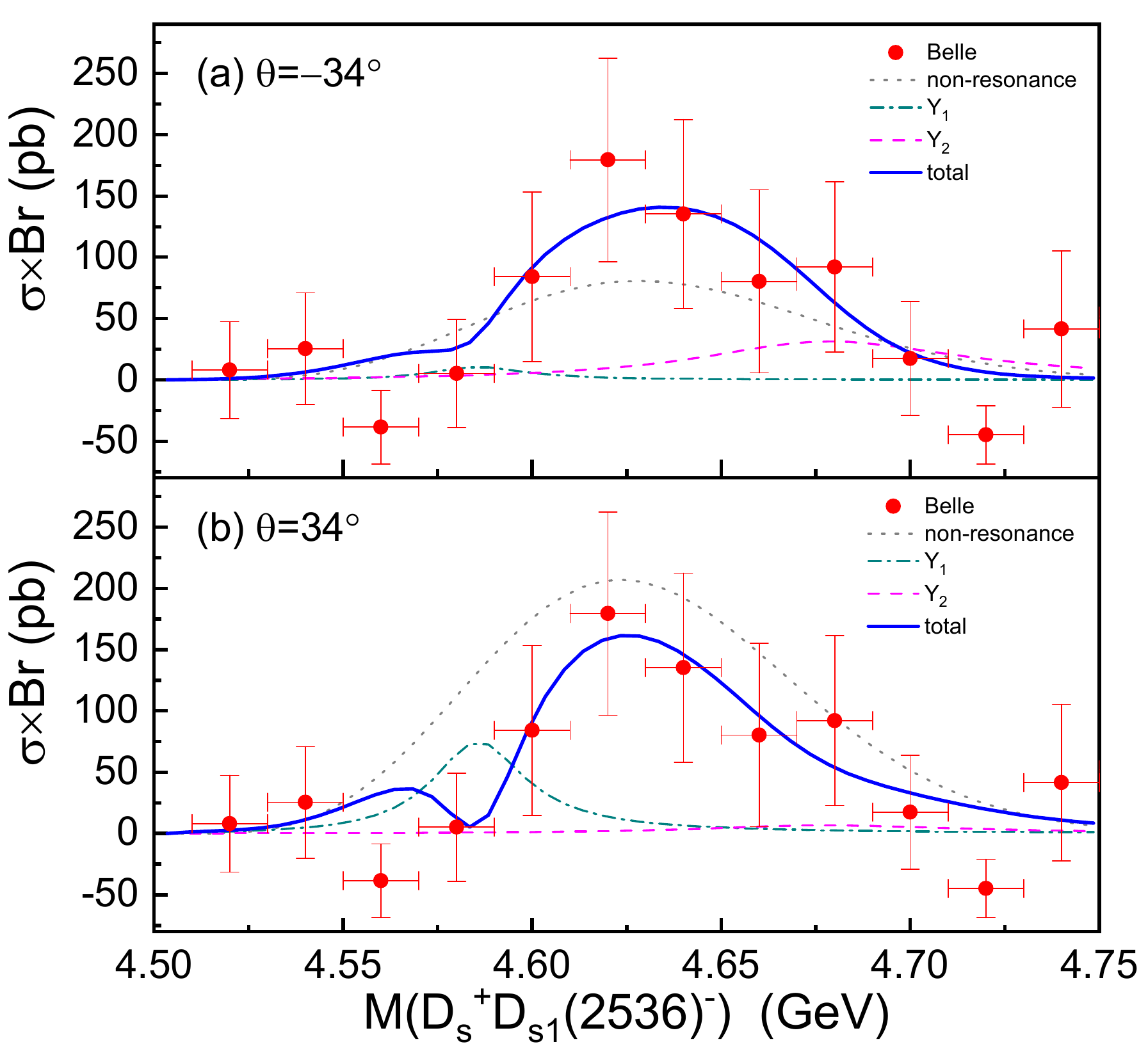}
	\caption{The reproduction of the experimental line shape of the $e^+e^-\to D_s^+ D_{s1}(2536)^-$ from the recent measurements of Belle \cite{Jia:2019gfe} in our two-resonance scheme.}
	\label{fig:dsds1exp}
	\end{figure}

With the above preparation, we fit the experimental data of the $e^+e^-\to D_s^+ D_{s1}(2536)^-$ by Belle in Fig.~\ref{fig:dsds1exp}, where the vertical coordinate stands for the product of cross section $\sigma(e^+e^-\to D_s^+ D_{s1}(2536)^-)$ and branching ratio $Br(D_{s1}(2536)^-\to \bar{D}^{*0} K^-)$. It can be seen that the experimental data of Belle can be reproduced very well in both the positive and negative angle scheme, which further enforce our conclusion that the observed $Y$ states around 4.6 GeV can be treated as higher charmonia. Additionally, the present experimental data from Belle is still rough and we expect more precise measurements which help us distinguish the sign of the mixing angle.

\section{Higher charmonia above 4.6 GeV \label{sec-higherccbar}}

Up to now, all of the observed charmoniumlike $Y$ states from $e^+e^-$ annihilation, including $Y(4220)$, $Y(4360)$, $Y(4630)$ and $Y(4660)$, can be reasonably assigned to the $J/\psi$ family in our proposed theoretical framework. In a sense, it indicates that the unquenched contributions have begun to challenge our knowledge inherent in the linear behavior of confinement potential in the charmonium sector. In this section, we will further predict the properties of higher charmonia above 4.6 GeV by the adopted unquenched potential model, where the screened parameter $\mu$= 0.12 can be fixed by scaling these $Y$ structures.

Here, we focus on six higher vector charmonia, which are $\psi(7S)$, $\psi(8S)$, $\psi(9S)$, $\psi(6D)$, $\psi(7D)$, and $\psi(8D)$. The numerical results of their masses and decay behaviors are listed in Table~\ref{tab:table1}. We can see that their masses are all located in the energy region of 4.70 to 4.90 GeV, which indicates that the charmonium spectroscopy above 4.6 GeV has become very dense.
Another interesting feature is that these high excited vector charmonia above 4.6 GeV are relatively narrow states although more open charm channels have opened. Their total widths are about 10 to 30 MeV as shown in Table ~\ref{tab:table1}. The main reason for this phenomenon is the node effect. In the QPC model, the decay amplitude is proportional to the overlap integration of the spatial wave function of initial and final states. As we all know, the meson radial wave function with the radial quantum number $n$ has $n$-1 nodes, which means the overlap integration of the highly radial excited charmonia may cancel each other in the positive and negative value area more strongly due to the existence of more nodes. Therefore, the interesting properties in the charmonium energy region above 4.6 GeV can be left to experimentalists to test in the future.

It needs to be emphasized that the bin size of present experimental data is not enough to identify higher charmonium states. {Fortunately, the energy region above 4.6 GeV can be accessed by the running BESIII and BelleII experiments. Very recently, we notice that the BESIII collaboration released their white paper on the future physics programme \cite{Ablikim:2019hff}, in which they plan to take data above 4.6 GeV to extend experimental search for more $Y$ states. They simply estimated number of events for BESIII and found it can be comparable to BelleII. However, a more important advantage of BESIII is that its energy resolution is relatively high \cite{Ablikim:2019hff}. It means smaller bin size of energy point can be achieved in BESIII, which is very significant to identify the possible narrow structures existing above 4.6 GeV. In addition to the running experiments, it is worth mentioning that the future Super Tau-Charm Factory as the successor of BEPCII has been discussed in China, which is planned to operate in
the range of center-of-mass energies from 2 to 7 GeV \cite{Luo:2019xqt}. Since the designed luminosity of Super Tau-Charm Factory is about 100 times larger than BEPCII, it will be an excellent platform to precisely study charmonium physics above 4.6 GeV. Based on the above experimental project, we strongly encourage experimentalists to pay more attention to the charmonium energy region above 4.6 GeV, where abundant physical phenomena and structures are expected to exist. This should be a good chance for BESIII and BelleII. }

According to our predictions of open charm decay behaviors listed in Table~\ref{tab:table1}, we can provide some dominant decay channels to search for the higher charmonia above 4.6 GeV in the future experiments. It can be seen that there are a plenty of open charm modes allowed by the phase space, where apart from two $S$-wave charmed mesons, the channels involving $S$-wave and $P$-wave charmed mesons are also important. For the higher $S$-wave charmonia above 4.6 GeV, we find the dominant decay channels are $DD_1(2430)$, $D^*D_0^*(2400)$, $D^*D_1(2430)$, $D^*D_2^*(2460)$, $DD^*(2600)$, and $D^*D_1(2420)$. However, the channels, $D^*D^*$, $DD$, $D^*D_1(2420)$, $D^*D_1(2430)$, $D^*D_2^*(2460)$, and $DD_1(2430)$ should be considered
for higher $D$-wave charmonia. {However some of these final
states like $D_1(2430)$, $D_0^*(2400)$ and $D^*(2600)$ will be quite broad, which makes it difficult if not impossible to
reconstruct the original state. 
Alternatively, three-body decay mode $DD^*\pi$ is recommended. Similarly, the mode of $D^*D^*\pi$ final states from decay channel $D^*D_1(2430)$ is suggested to search for the charmonium states.} In addition to the above dominant charmed meson modes, the present experimental measurements have proved that the charmed-strange meson, hidden-charm and charmed baryon channels can also be measurable modes. In short, we are very much looking forward to precise measurements of these decay processes above 4.6 GeV in the future BESIII and BelleII, {and even in the possible Super Tau-Charm Factory,} although it may be a great challenge to experimentalists.

	\begin{table*}
	\caption{\label{tab:table1}%
	   The decay behaviors of higher charmonia with $J^{PC}=1^{--}$ without considering $S$-$D$ mixing, where {higher charmed and charmed-strange mesons predicted in the potential model } are labeled by $n^{2S+1}L_J$. 
	   {For the notation of charmonium}, we omit $2S+1$ and $J$ because they are 3 and 1, respectively. The dots represent that the channel is forbidden by either quantum number or phase space. The mass spectrum is given with $\mu$=0.12 and all of the results are in units of MeV.}
	\begin{ruledtabular}
	\begin{tabular}{ccccccccc}
			\textrm{}&\textrm{$\psi(6S)$}&\textrm{$\psi(7S)$}&\textrm{$\psi(8S)$}&\textrm{$\psi(9S)$}&\textrm{$\psi(5D)$}&\textrm{$\psi(6D)$}&\textrm{$\psi(7D)$}&\textrm{$\psi(8D)$}\\
			\colrule
			\textrm{Mass} & 4615 & 4726 & 4808 & 4867 & 4648 & 4750 & 4826 & 4880\\
			\textrm{Total width} & 28.50 & 27.60 & 23.11 & 17.07 & 27.35 & 19.77 & 14.71 & 10.19 \\
			\hline
			\textrm{Channel} &  &  &  &  &  &  &  & \\
			\textrm{$DD$} & 1.49 & 1.13 & 0.81 & 0.55 & 4.33 & 2.98 & 2.04 & 1.31\\
			\textrm{$DD^{*}$} & 0.40 & 0.49 & 0.45 & 0.35 & 0.76 & 0.58 & 0.44 & 0.28\\
			\textrm{$D^{*}D^{*}$} & 3.06 & 1.29 & 0.60 & 0.29 & 4.22 & 3.17 & 2.32 & 1.52\\
			\textrm{$DD_0^*(2400)$} & $\dotsb$ & $\dotsb$ & $\dotsb$ & $\dotsb$ & $\dotsb$ & $\dotsb$ & $\dotsb$ & $\dotsb$\\
			\textrm{$DD_1(2420)$} & 7.09 & 5.41 & 3.80 & 2.51 & 2.04 & 1.09 & 0.62 & 0.36\\
			\textrm{$DD_1(2430)$} & 1.85 & 0.92 & 0.53 & 0.32 & 2.76 & 1.04 & 0.50 & 0.29\\
			\textrm{$DD_2^*(2460)$} & 2.23 & 1.10 & 0.51 & 0.23 & 0.53 & 0.14 & 0.03 & 0.01\\
			\textrm{$D^{*}D_0^*(2400)$} & 5.89 & 5.27 & 3.94 & 2.69 & 1.66 & 1.01 & 0.63 & 0.38\\
			\textrm{$DD(2550)$} & 1.33 & 0.42 & 0.10 & 0.02 & 1.18 & 0.11 & $10^{-3}$ & 0.02\\
			\textrm{$DD^{*}(2600)$} & 1.44 & 2.05 & 1.29 & 0.69 & 1.75 & 1.03 & 0.49 & 0.23\\
			\textrm{$D^{*}D_1(2420)$} & 1.14 & 3.34 & 3.60 & 2.92 & 1.72 & 2.01 & 1.66 & 1.18\\
			\textrm{$D^{*}D_1(2430)$} & 1.09 & 2.19 & 1.90 & 1.32 & 3.82 & 3.33 & 2.19 & 1.31\\
			\textrm{$D^{*}D_2^*(2460)$} & 0.02 & 2.01 & 2.70 & 2.23 & 1.50 & 1.35 & 0.86 & 0.50\\
			\textrm{$D^{*}D(2550)$} & 0.05 & 0.58 & 0.89 & 0.70 & 0.22 & 0.78 & 0.62 & 0.38\\
			\textrm{$D^{*}D^{*}(2600)$} & $\dotsb$ & 0.02 & 0.44 & 0.95 & 0.08 & 0.36 & 0.88 & 0.74\\
			\textrm{$DD(^{3}D_{1})$} & $\dotsb$  & $10^{-5}$ & 0.02 & 0.03 & $10^{-4}$ & 0.02 & 0.06 & 0.06\\
			\textrm{$DD(^{1}D_{2})$} & $\dotsb$  & $10^{-3}$ & 0.03 & 0.07 & 0.01 & 0.04 & 0.34 & 0.38\\
			\textrm{$DD(^{3}D_{2})$} & $\dotsb$ & $10^{-3}$ & 0.02 & 0.06 & $10^{-3}$ & 0.01 & 0.09 & 0.11\\
			\textrm{$DD^*(2760)$} & $\dotsb$ & $10^{-3}$ & 0.05 & 0.08 & $10^{-3}$ & 0.02 & 0.06 & 0.05\\
			\textrm{$D_0^*(2400)D_0^*(2400)$} & $\dotsb$ & $10^{-6}$ & 0.01 & 0.02 & $10^{-3}$ & 0.03 & 0.11 & 0.13\\
			\textrm{$DD(2^{3}P_{0})$} & $\dotsb$ & $\dotsb$ & $\dotsb$ & $\dotsb$ & $\dotsb$  & $\dotsb$ & $\dotsb$ & $\dotsb$\\
			\textrm{$DD(2^{1}P_{1})$} & $\dotsb$ & $10^{-4}$ & 0.02 & 0.09 & $\dotsb$ & 0.03 & 0.22 & 0.45\\
			\textrm{$DD(2^{3}P_{1})$} & $\dotsb$ & $\dotsb$ & 0.04 & 0.11 & $\dotsb$ & 0.01 & 0.06 & 0.14\\
			\textrm{$DD(2^{3}P_{2})$} & $\dotsb$ & $\dotsb$ & $10^{-4}$ & $10^{-3}$ & $\dotsb$ & $\dotsb$ & $10^{-3}$ & 0.02\\
			\textrm{$D_{s}D_{s}$} & $10^{-3}$ & $10^{-3}$ & $10^{-3}$ & $10^{-3}$ & 0.01 & 0.01 & 0.01 & 0.01\\
			\textrm{$D_{s}D_{s}^{*}$} & 0.12 & 0.06 & 0.03 & 0.01 & 0.07 & 0.04 & 0.02 & 0.01\\
			\textrm{$D_{s}^{*}D_{s}^{*}$} & 0.31 & 0.22 & 0.14 & 0.09 & 0.09 & 0.05 & 0.03 & 0.02\\
			\textrm{$D_{s}D_{s0}^*(2317)$} & $\dotsb$ & $\dotsb$ & $\dotsb$ & $\dotsb$ & $\dotsb$ & $\dotsb$ & $\dotsb$ & $\dotsb$\\
			\textrm{$D_{s}^{*}D_{s0}^*(2317)$} & 0.8 & 0.77 & 0.84 & 0.41 & 0.23 & 0.15 & 0.09 & 0.06\\
			\textrm{$D_{s}D_{s1}(2460)$} & 0.01 & 0.03 & 0.03 & 0.02 & 0.21 & 0.18 & 0.13 & 0.08\\
			\textrm{$D_{s}^{*}D_{s1}(2460)$} & 0.01 & 0.01 & 0.01 & 0.05 & $10^{-3}$ & 0.01 & 0.02 & 0.02\\
			\textrm{$D_{s}D_{s2}^*(2573)$} & $10^{-4}$ & $10^{-3}$ & 0.01 & 0.01 & $10^{-3}$ & 0.01 & 0.01 & 0.01\\
			\textrm{$D_{s}^{*}D_{s2}^*(2573)$} & $\dotsb$ & $10^{-5}$ & $10^{-4}$ & $10^{-6}$ & $\dotsb$ & $10^{-4}$ & $10^{-4}$ & $10^{-3}$\\
			\textrm{$D_{s}D_{s1}(2536)$} & 0.17 & 0.23 & 0.21 & 0.16 & 0.16 & 0.13 & 0.10 & 0.06\\
			\textrm{$D_{s}^{*}D_{s1}(2536)$} & $\dotsb$ & 0.06 & 0.08 & 0.07 & $10^{-4}$ & 0.01 & 0.02 & 0.02\\
		\end{tabular}
\end{ruledtabular}
    \end{table*}

\section{SUMMARY\label{sec-summary}}

In the past decades, the studies of charmoniumlike $Y$ states from the $e^+e^-$ annihilation have become a very hot subject. Since the first observation of the super star $Y(4260)$, more and more $Y$ structures were reported by experiments, which form a special group in the present $XYZ$ family as summarized in Refs. \cite{Chen:2016qju,Liu:2019zoy}. Although their properties have brought great challenges to the conventional charmonium spectrum, it is still a good opportunity to reveal some new features of highly excited $c\bar{c}$ mesons. In the previous work \cite{Wang:2019mhs}, the charmoniumlike state $Y(4220)$ has been established as a charmonium under $4S$-$3D$ mixing scheme, and further studies have indicated that a charmonium partner $\psi(4380)$ of $Y(4220)$ should exist. Thus, all of the charmoniumlike $Y$ states below 4.5 GeV have been well explained in our proposed unquenched framework. However, the nature of the remaining two structures $Y(4630)$ and $Y(4660)$ is still a mystery.

In this work, we have studied the mass spectra and open charm decay behaviors of higher charmonia by using the same approach as Ref. \cite{Wang:2019mhs}. Then, we have further discussed whether the charmoniumlike $Y$ states around 4.6 GeV can be treated as higher charmonia. We have found it is not suitable to assign $Y(4660)$ observed in the $e^+e^-\to \psi(2S)\pi^+\pi^-$ to a charmonium $\psi(6S)$ due to the mass inconsistence. By analyzing a combined experimental data for the $e^+e^-\to \psi(2S)\pi^+\pi^-$ and $e^+e^- \to \Lambda_c \bar{\Lambda}_c$, where the recent measurements from BESIII are included, we have pointed out that there may exist another potential resonance with lower mass besides two known $Y(4630)$ and $Y(4660)$. The center mass and width of this underlying structure is fitted to 4585 and 29.8 MeV, respectively, which are exactly consistent with our predicted charmonium $\psi^{\prime}_{6S-5D}$ in the $6S$-$5D$ mixing scheme. Furthermore, the $\psi^{\prime\prime}_{6S-5D}$ with main component of a $5^3D_1$ $c\bar{c}$ state can correspond to the reported $Y(4630)$ and $Y(4660)$ states, where these two charmoniumlike structures could be from the same source in our fit. Therefore, our theoretical results presented here support that the charmoniumlike $Y$ states around 4.6 GeV are explained to be higher charmonia. Finally, we have also given the predictions of masses and decay behaviors of higher vector charmonia above 4.6 GeV.

Very recently, the Belle reported the observed results of an open charm process $e^+e^-\to D_s^+ D_{s1}(2536)^-+c.c.$ \cite{Jia:2019gfe}, where a peak at 4.63 GeV is found. This is the first experimental discovery of the $Y$ structure around 4.6 GeV in an open charm channel. We have further explored partial decay widths of two charmonium states $\psi^{\prime}_{6S-5D}$ and $\psi^{\prime\prime}_{6S-5D}$ for the $D_s^+ D_{s1}(2536)^-+c.c.$ mode and have found the experimental data of Belle can be well reproduced both in negative and positive mixing angle schemes. This strong evidence from an open charm channel further strengthen our confidence for the nature of the charmoniumlike $Y$ states around 4.6 GeV as $c\bar{c}$ states.

Our theoretical research has provided some new insights into the high excited charmonium spectrum, where all of the observed charmoniumlike $Y$ states can be well settled down by the $J/\psi$ family without introducing any exotic configurations. There is no doubt that the charmonium physics between 4.5 to 5.0 GeV will be very interesting and the studies on them can greatly improve our understanding for the confinement interaction in $c\bar{c}$ system. We hope that more experimental and theoretical groups could focus on this topic in the future.

\section*{Acknowledgement}

This work is supported in part by the China National Funds for Distinguished Young Scientists under Grant No. 11825503 and the National Program for Support of Top-notch Young Professionals.


\end{document}